\documentclass[3p]{elsarticle}

\usepackage[utf8]{inputenc}
\usepackage[T1]{fontenc}
\usepackage{newtxtext} 
\usepackage{amsmath,amssymb,amsfonts,amsthm}
\usepackage{mathtools}
\usepackage{algorithm}
\usepackage{algpseudocode}
\makeatletter
\providecommand*{\theHALG@line}{\thealgorithm.\arabic{ALG@line}}
\makeatother
\usepackage{booktabs}
\usepackage{multirow}
\usepackage{tabularx}
\usepackage{xltabular}
\usepackage{array}
\usepackage{float}
\usepackage{placeins}
\usepackage{eurosym}
\allowdisplaybreaks
\usepackage{graphicx}
\usepackage{setspace}
\usepackage{xcolor}
\usepackage[normalem]{ulem}
\usepackage{geometry}
\usepackage{hyperref}
\usepackage{enumitem}
\hypersetup{colorlinks=true, breaklinks=true, citecolor=teal, linkcolor=teal, urlcolor=blue}
\usepackage{cleveref}

\usepackage{tikz}
\usetikzlibrary{arrows.meta, positioning, shapes.geometric, calc, fit, backgrounds}
\usepackage{orcidlink} 

\definecolor{navy}{RGB}{31,97,141}
\definecolor{teal}{RGB}{22,160,133}
\definecolor{coral}{RGB}{230,126,34}
\definecolor{crimson}{RGB}{203,67,53}
\definecolor{amber}{RGB}{243,156,18}
\definecolor{forest}{RGB}{39,174,96}
\definecolor{slate}{RGB}{128,139,150}

\newtheorem{prop}{Proposition}

\newtheorem{ass}{Assumption}
\newtheorem{remark}{Remark}
\newtheorem{definition}{Definition}
\crefname{prop}{Proposition}{Propositions}
\Crefname{prop}{Proposition}{Propositions}
\crefname{ass}{Assumption}{Assumptions}
\Crefname{ass}{Assumption}{Assumptions}
\crefname{alg}{Algorithm}{Algorithms}
\newtheorem{prof}{Proof}

\newcommand{\R}{\mathbb{R}}

\usetikzlibrary{arrows.meta,shapes.geometric,positioning,calc,backgrounds,fit}
\definecolor{tabblue}{HTML}{1F77B4} \definecolor{taborange}{HTML}{FF7F0E}
\definecolor{tabgreen}{HTML}{2CA02C} \definecolor{tabred}{HTML}{D62728}
\journal{Applied Energy}

\begin{document}
	\begin{frontmatter}
		
		\title{What the annual energy balance hides: the coordination value of digital platforms for positive energy districts}
		
		\author[dtu]{Arega Getaneh Abate\orcidlink{0000-0002-5517-2585}\corref{cor1}}
		\ead{ageab@dtu.dk}
		\author[dtu]{Per Sieverts Nielsen\orcidlink{0000-0003-4090-7400}}
		\ead{pernn@dtu.dk}
		\author[dtu]{Xiufeng Liu\orcidlink{0000-0001-5133-6688}}
		\author[ucc]{Brian O'Regan}
		\author[eff]{Gerfried Cebrat}
		\author[tub]{Cigdem Karadag}
		\author[mdu]{Valentina Zaccaria}
		\author[liu]{Yang Liu}
		\author[beia]{Madalin Silion}
		\author[mdu]{Amir Vadiee}
		\author[ltk]{Jakob Jespersen}
		\author[inn]{Mahmut Sami B\"uker}
		\author[icsi]{Maria Simona Raboaca}
		\author[mdu]{Avinash Renuke}
		\author[tub]{Bet\"ul Erd\"or T\"urk}
		\author[mega]{Dudley Stewart}
		\cortext[cor1]{Corresponding author.}
		\affiliation[dtu]{organization={Technical University of Denmark},country={Denmark}}
		\affiliation[ucc]{organization={Tyndall National Institute, University College Cork},country={Ireland}}
		\affiliation[eff]{organization={effiziente.st},country={Austria}}
		\affiliation[tub]{organization={TÜBİTAK MAM},country={Türkiye}}
		\affiliation[mdu]{organization={Mälardalen University},country={Sweden}}
		\affiliation[liu]{organization={Linköping University},country={Sweden}}
		\affiliation[beia]{organization={BEIA Consult International},country={Romania}}
		\affiliation[ltk]{organization={Lyngby-Taarbæk Municipality},country={Denmark}}
		\affiliation[inn]{organization={Innorma \& Necmettin Erbakan University},country={Türkiye}}
		\affiliation[icsi]{organization={ICSI},country={Romania}}
		\affiliation[mega]{organization={MEGA},country={Ireland}}
		
		\begin{abstract}
	Positive energy districts (PEDs) are usually assessed by annual net energy, whereas distribution networks operate hour by hour. An annually positive district still imports on low-generation hours and exports or curtails on high-generation hours. How large this hourly imbalance is behind an annual balance, and how much of it coordination among prosumers can close without new physical capacity, is not established. This paper models a multi-carrier prosumer community coupling electricity and heat behind a shared grid connection with a net-export hosting limit. The model is solved as a centralized community problem and as a price-taking peer-to-peer (P2P) market. Under convexity, the equilibrium implements the regularized \textit{community optimum} and the \textit{P2P and congestion} prices are the duals of the pool-clearing and hosting constraints. The model is calibrated to six European district archetypes with constructed hourly profiles. Hourly self-sufficiency is only 48--50\% in the four deeply studied districts and ranges from 48--55\% across all six, leaving a no-platform coordination gap of 2.8--5.9\% of no-platform economic cost. The shared-forecast step recovers only 6--18\% of the gap, while grid-aware pricing reduces curtailment by 45\% in the primary district. Coordination mainly reduces surplus-side inefficiency, whereas closing the deficit side requires storage. The results are scenario-based projections for calibrated archetypes, and the operating benefit is small and contingent on binding hosting limits and high feed-in remuneration. For regulators and distribution system operators this supports hourly, bidirectional indicators alongside the annual positive-energy target, and a congestion charge active only in binding hours with a predetermined refund, while the business case for a platform at district scale rests on grid-side value and aggregation across districts.
		\end{abstract}
		
		
	\end{frontmatter}
	
	\section{Introduction}\label{sec_introduction}
	
	A positive energy building produces more renewable energy on site than it consumes over a year. At the district scale, a positive energy district (PED) is a connected group of buildings that meets the same test on an annual basis \citep{brozovsky2021zero,sareen2022ten}. The concept has become a central instrument of European urban decarbonization. A substantial literature has refined its definition, its design methodologies, and its assessment frameworks \citep{sassenou2024methodologies}. Definitions vary, and recent work introduces virtual, dynamic, and autonomy-oriented variants \citep{nikolic2025letsmake}, although many PED assessments still operationalize positivity through an annual net balance. A district is declared positive when its renewable generation over a year exceeds its demand over the same year. This is distinct from an hourly matching target, under which supply must match demand in every hour. The annual test can be met with large hourly mismatches that an hourly target would penalize.
	
	An annual balance is a useful decarbonization target. However, it conceals the operational problem that matters to the distribution grid. A district that is annually positive may still import heavily on winter evenings and export or spill generation at summer middays. This temporal mismatch does not disappear once annual balance is met, and it is what the distribution system operator (DSO) experiences hour by hour. A related literature on hourly, or temporal, matching of clean electricity, sometimes termed 24/7 carbon-free energy, makes the same point for corporate procurement. It shows that hourly matching is a far stronger requirement than annual matching, and that it alters both the cost and the technology mix \citep{xu2024systemlevel,riepin2024means}. The district counterpart of this problem has received less attention. The annual mismatch is bidirectional, which makes the problem non-trivial to address. Undersupply obliges the community to import, whereas oversupply obliges it to export or to curtail surplus generation. Oversupply can also create distinct network effect and market costs, through reverse power flows, voltage rise, curtailment, and exposure to negative prices, that an annual net-energy indicator does not capture. Therefore, a district that oversupplies at midday and undersupplies at night imposes two distinct demands on the network, and an annual balance measures neither. We call this bidirectional residual demand and supply the \emph{residual hourly imbalance}. In the positive energy districts studied in this paper, the residual hourly imbalance reaches 45--52\% of demand on the deficit side and 46--53\% of renewable output on the surplus side. As a result, the engineering and economic difficulty lies in closing it without shifting the burden from deficit to surplus.
	
	A second challenge in PED analysis is how the residual gap is closed, and by which organization the challenge can be addressed as a PED is not a single optimizer. It is a community of prosumers (energy buyers and sellers), each with its own assets, preferences, and information, connected behind a common grid coupling and interacting with a DSO. Such a community can be organized in different ways \citep{cebrat2026ecellular}, ranging from a central operator that co-optimizes every asset under full information to fully decentralized prosumers that act independently \citep{sousa2019peer}, trade peer-to-peer (P2P) \citep{tushar2021peer}, and respond to prices. The difference in outcome between these two organizations of the same physical system, the central operator and the decentralized community, is an economic quantity that \textit{interoperable digital platforms} are intended to reduce. Data exchange, digital twins, shared forecasts, and market mechanisms are commonly justified by the claim that they let a decentralized community approach the performance of a coordinated one. We call this difference the coordination gap. Its size, and how much of it each platform mechanism closes, is the second question this paper answers.
	
	Three structural features make the problem demanding. First, the district is a multi-carrier system in which electricity and heat are coupled by heat pumps and thermal storage. The balancing of one carrier cannot be judged in isolation from the other. Second, the resources that close the gap, namely batteries, thermal storage, and shiftable loads, are intertemporal. Therefore, the value of these balancing resources depends on foresight and on coordination across agents and across time. Third, the coordination itself is an economic object with a value of its own. The same physical district yields different outcomes under central control, under a P2P market, and under different grid tariff designs.
	
	Closing this research gap motivates three research questions. \textit{First, can one tractable model represent the multi-carrier district, the P2P market among its prosumers, and the DSO hosting constraint jointly, while keeping the coordination signals interpretable? Second, how large is the coordination gap between independent operation and a centralized community operator, and what ordered marginal contributions come from shared forecasts, P2P trading, and grid-aware pricing? Third, how much of the residual hourly gap can operational coordination reduce, and how much remains bounded by installed storage and therefore requires investment?}
	
	To address these research questions, we develop a multi-carrier community optimization model and calibrate the model to six European district configurations. We solve the same physical system as a centralized community problem and as a decentralized P2P market equilibrium. Under convexity, the equilibrium implements the regularized community optimum. The economic P2P and congestion prices are obtained from the dual multipliers of the market-clearing and hosting constraints. This duality makes the equilibrium computable, the prices interpretable, and the organizational comparison transparent. Every value we present is a scenario-based projection for calibrated district archetypes, and the interpretations and implications are drawn within that scope.
	
	This paper integrates elements that existing literature treats separately. One convex model represents the multi-carrier district, the P2P pool among its prosumers, and the net-export hosting limit at the shared coupling point. The market and congestion prices are derived from the dual multipliers of the two coupling constraints of that model, which makes them shadow values of stated physical constraints instead of assumed tariffs. The same duality then lets the decentralized equilibrium and the centralized operator be compared on identical physical ground. An ordered attribution separates what shared forecasts deliver from what organization and grid pricing deliver, and keeps the one forecast-informed stage distinct from the perfect-information stages. A bidirectional hourly metric splits the residual into a deficit share and an oversupply share, which shows that annual positivity is consistent with a large and asymmetric hourly imbalance. The paper covers six calibrated European district archetypes and eight sensitivity dimensions, and throughout the analysis separates the value operational coordination delivers from the part that installed storage must provide.
	
	The remainder of the paper is organized as follows. \Cref{sec_literature} reviews the related work and positions the contribution of this paper. \Cref{sec_methodology} presents the multi-carrier model, the two coordination paradigms, the duality that connects them, and the solution method. \Cref{sec_setup} describes the testbed, the data calibration, and the experimental design. \Cref{sec_results} presents the results and the sensitivity analysis, and finally, \Cref{sec_conclusion} concludes the paper.
	
	\section{Related work}\label{sec_literature}
	
	The problem considered in this paper sits at the intersection of four strands of the literature, joined by a fifth methodological mix. Positive energy district literature establishes the annual energy concept while leaving operational coordination as given. Temporal-matching literature argues that hourly balance is the demanding target, particularly at the system level. Flexibility literature characterizes the physical resource that closes the gap without the market that dispatches it. Peer-to-peer market literature designs the market, though rarely connecting it to a physically explicit multi-carrier district. A methodological literature on distributed optimization and market equilibrium bridges central and decentralized organization, though rarely applied to the bidirectional hourly balance of a PED. In the following, we review the closest work in each strand and position the current paper.
	
	\citet{sareen2022ten} consolidate the PED concept in ten questions on definitions, boundary conditions, and assessment, and \citet{sassenou2024challenges} review the challenges that deployment has met. More specifically, \citet{nikolic2025letsmake} argue that the annual energy balance is only a first-order indicator and that temporal resolution, grid interaction, and flexibility are needed to judge whether a district is beneficial to the energy system.  \citet{bossi2020towards} analyze 61 European cases and document the diversity of sites and the gap between the annual target and operational performance. These studies establish the concept and the design space. However, they evaluate the district under a fixed operational framework and do not treat the coordination of its assets, and the market among its prosumers as a decision.
	
	The debate on temporal matching of clean electricity has made the quantitative case that hourly matching is far more demanding than annual matching. \citet{xu2024systemlevel} show for corporate procurement that hourly matching changes the cost and the technology mix, with very different system-level effects depending on how it is pursued. \citet{riepin2025matching} argue that such matching accelerates the adoption of advanced clean technologies, and \citet{riepin2024spatiotemporal} extend the same approach to flexible demand through spatio-temporal load shifting. This literature works at the level of a firm or a system, not a distribution-connected community. It does not treat the bidirectional grid burden or the DSO hosting limit that define the district problem. This paper extends this temporal argument to the district scale and treats it as bidirectional.
	
	Energy flexibility is the physical resource that closes the hourly gap. \citet{reynders2017generic} give a generic method to characterize the flexibility of structural thermal storage in residential buildings, and \citet{li2021energy} review the characterization and quantification methods for building flexibility across thermal mass, thermal storage, and controllable loads. This motivates the representations of electrical and thermal storage with realistic efficiencies and losses, and of bounded shiftable demand as a virtual store, that we use. The base model takes building heat demand as exogenous through the calibrated degree-hour model, and a sensitivity adds a bounded fabric store.
	
	\citet{sousa2019peer} and \citet{tushar2021peer} review peer-to-peer and community-based markets as the market layer of prosumer communities and catalog the market designs proposed for them. \citet{sorin2019consensus} clear a P2P market by consensus with product differentiation, and \citet{goitiazabaleta2023twostage}  manage a community-based P2P market in two stages with a central coordinator. The strand closest to this paper connects P2P trading to the physical distribution network. \citet{guerrero2019decentralized} trade peer-to-peer under network constraints in a low-voltage feeder, and \citet{jin2020local} review local flexibility markets, in which network needs are priced and allocated to trades. In this paper, we adopt a community pool with a uniform clearing price and add a DSO congestion price, where we connect both to the planner through duality. This enables us to drive the price instead of taking it by assumption.
	
	The relationship between centralized optimization and decentralized markets is a classical theme of energy economics. \citet{lecadre2020peer} analyze P2P market equilibria from the variational to the generalized Nash setting.  \citet{kraning2014dynamic} supply the algorithmic bridge between central and decentralized solutions through the alternating direction method of multipliers (ADMM) and proximal message passing. We compute the equilibrium as the convex community problem and take the market and congestion prices from its duals, prices that local price adjustment could in principle reach without our developing that iteration. Districts that couple electricity and heat also require a multi-energy treatment. \citet{geidl2007optimal} formulate the energy hub, in which carriers are linked by conversion and storage, and \citet{ramsebner2021sector} review the sector-coupling concept built on it. Finally, digital twins have been proposed as the information backbone of energy systems, supplying calibrated models, state estimation, and fast what-if analysis \citep{doamaral2023energy,agostinelli2021cyber}. The shared-forecasting layer adopted in this paper is a decision-support platform and not a full digital twin. It distributes common day-ahead forecasts without state estimation, synchronization, or model updating. Blockchain and transactive-energy concepts provide the settlement layer for P2P trading \citep{andoni2019blockchain,abrishambaf2019towards}, usually justified qualitatively.
	
	Several recent studies approach the joint problem and help position the contribution of this paper. \citet{wang2022peer} design a multi-energy P2P market with cooperative behavior, and \citet{li2025digital} pair a digital twin with P2P sharing across multiple energy hubs, though neither imposes a distribution-grid hosting limit nor connects the market prices to a planner benchmark. \citet{sun2023network} make P2P trading network-aware in a combined electricity-heat system, and \citet{bjarghov2024enhancing} raise hosting capacity through coordinated non-firm connections, both bringing the grid explicitly into the community problem, while lacking the bidirectional hourly self-sufficiency framing and the planner-equilibrium duality. \citet{taxt2025integration} map development paths for integrating communities into distribution grids. \citet{crowley2025how} address budget balance, individual rationality, fair allocation, and dynamic grid-service pricing, the settlement and participation properties that we treat only in reduced form and leave to future work. Relative to this body of work, the main contribution of this paper is an integrated convex model of a multi-carrier district, its P2P pool, and the net-export hosting limit at the coupling point, in which the market and congestion prices are derived as the multipliers of the two coupling constraints. The same duality makes the decentralized equilibrium and the central operator comparable on identical physical ground. The coordination gap is decomposed into three mechanisms, shared forecasts, P2P trading, and grid-aware pricing, and each is valued on calibrated district archetypes. The decomposition enables to show that operational coordination acts on the surplus side of the hourly imbalance and that the deficit side is bounded by installed storage. Table \ref{tab_positioning} compares the paper with the closest studies across six features, where \emph{no} marks a feature absent or exogenous, \emph{partial} one treated in reduced form, and \emph{yes} one that is central.
	
	\begin{table}[H]
		\centering
		\caption{Feature comparison with the closest recent studies.}
		\label{tab_positioning}
		\setlength{\tabcolsep}{4pt}
		\begin{tabularx}{\textwidth}{@{}>{\raggedright\arraybackslash}p{3.1cm}*{6}{>{\centering\arraybackslash}X}@{}}
			\toprule
			Study & Multi-carrier & Hourly balance & Export hosting & P2P market & Planner duality & Grid price \\
			\midrule
			\citet{wang2022peer}        & yes     & partial & no      & yes & partial & no \\
			\citet{sun2023network}      & yes     & partial & yes     & yes & partial & partial \\
			\citet{bjarghov2024enhancing} & no    & partial & yes     & no  & no      & partial \\
			\citet{li2025digital}       & yes     & partial & no      & yes & no      & no \\
			\citet{crowley2025how}      & no      & partial & partial & yes & partial & yes \\
			\citet{taxt2025integration} & partial & no      & partial & no  & no      & partial \\
			\text{The proposed model}         & \text{yes} & \text{yes} & \text{yes} & \text{yes} & \text{yes} & \text{yes} \\
			\bottomrule
		\end{tabularx}
	\end{table}

	\section{Methodology}\label{sec_methodology}
	
	\subsection{Problem setting and community representation}\label{subsec_setting}
	
\begin{figure}[htbp]
		\centering
		\resizebox{1\textwidth}{!}{%
			\begin{tikzpicture}[font=\rmfamily,
				ag/.style={rectangle, draw=tabblue, fill=tabblue!8, line width=0.7pt, rounded corners=3pt, text width=3.7cm, minimum height=0.9cm, font=\small, text=black, align=left, inner xsep=7pt, inner ysep=4pt},
				pool/.style={rectangle, draw=tabgreen!80!black, fill=tabgreen!10, line width=1.0pt, rounded corners=5pt, minimum width=2.6cm, minimum height=5.3cm, text=black, align=center, font=\small},
				pcc/.style={circle, draw=tabred!85!black, fill=tabred!8, line width=0.9pt, minimum size=1.45cm, font=\small\bfseries, text=black, align=center},
				grid/.style={rectangle, draw=tabred!85!black, fill=tabred!8, line width=0.8pt, rounded corners=3pt, text width=3.6cm, minimum height=2.1cm, font=\small, text=black, align=center, inner ysep=5pt},
				twin/.style={rectangle, draw=taborange!90!black, fill=taborange!10, line width=0.8pt, dashed, rounded corners=3pt, minimum width=8.6cm, minimum height=1.0cm, font=\small, text=black, align=center},
				flow/.style={-{Stealth[length=2.5mm]}, line width=0.8pt, tabred!85!black},
				flowbd/.style={{Stealth[length=2.5mm]}-{Stealth[length=2.5mm]}, line width=0.9pt, black},
				trade/.style={{Stealth[length=2.2mm]}-{Stealth[length=2.2mm]}, line width=0.8pt, tabgreen!80!black},
				twinarr/.style={-{Stealth[length=2mm]}, taborange!90!black, line width=0.7pt, dashed},
				header/.style={font=\footnotesize\scshape, text=black},
				lab/.style={font=\scriptsize, text=black, fill=white, inner sep=1.5pt}]
				
				\node[header] at (0,4.15) {Prosumer agents};
				\node[header] at (5.2,4.15) {Community market};
				\node[header] at (9.6,4.15) {Coupling point};
				\node[header] at (14.6,4.15) {Distribution grid};
				
				\node[ag] (a1) at (0,2.45) {\textbf{Residential block}\\[-1pt]{\scriptsize PV, battery, heat pump}};
				\node[ag] (a2) at (0,1.05)  {\textbf{Residential block}\\[-1pt]{\scriptsize PV, thermal store, DH}};
				\node[ag] (a3) at (0,-0.35){\textbf{Public building}\\[-1pt]{\scriptsize PV/T, heat pump, flexible load}};
				\node[ag] (a4) at (0,-1.75) {\textbf{Office}\\[-1pt]{\scriptsize PV facade, HVAC}};
				
				\node[pool] (pool) at (5.2,0.35) {\textbf{P2P pool}\\[3pt]{\footnotesize uniform clearing}\\{\footnotesize price $\pi_t$}\\[5pt]{\footnotesize offers balance bids}};
				
				\node[pcc] (pcc) at (9.6,0.35) {PCC};
				\node[grid] (grid) at (14.6,0.35) {\textbf{Grid / DSO}\\[3pt]{\footnotesize hosting limit $\bar E_t$}\\{\footnotesize congestion price $\tau_t$}\\{\footnotesize tariffs $\lambda^{\mathrm{imp}}_t,\ \lambda^{\mathrm{exp}}_t$}};
				
				\foreach \a in {a1,a2,a3,a4}{ \draw[trade] (\a.east) -- (\a.east -| pool.west); }
				\node[lab] at ($(a1.east)!0.5!(a1.east -| pool.west)+(0,0.25)$) {buy / sell};
				
				\draw[flowbd] (pool.east) -- (pcc.west) node[midway, above, lab, yshift=2pt] {net exchange};
				
				\draw[flow] ([yshift=3.2mm]grid.west) -- ([yshift=3.2mm]pcc.east);
				\node[lab] at ($([yshift=3.2mm]pcc.east)!0.5!([yshift=3.2mm]grid.west)+(0,0.3)$) {import $I_t$};
				\draw[flow] ([yshift=-3.2mm]pcc.east) -- ([yshift=-3.2mm]grid.west);
				\node[lab] at ($([yshift=-3.2mm]pcc.east)!0.5!([yshift=-3.2mm]grid.west)+(0,-0.3)$) {export $E_t\le\bar E_t$};
				
				\begin{scope}[on background layer]
					\node[draw=black, rounded corners=7pt, inner xsep=9pt, inner ysep=9pt, fit=(a1)(a4)(pool), line width=0.5pt, draw=tabblue!60, densely dotted] (comm) {};
				\end{scope}
				\node[anchor=south west, font=\scriptsize\itshape] at ($(comm.north west)+(0.05,0.03)$) {energy community};
				
				\node[twin] (twin) at (2.6,-4.0) {\textbf{Shared-forecasting platform}\quad{\footnotesize common day-ahead forecasts of load, PV, and prices}};
				\draw[twinarr] (twin.north -| a3) -- (comm.south -| a3);
				\draw[twinarr] (twin.north -| pool) -- (comm.south -| pool);
				\node[lab, anchor=west] at ($(twin.north -| pool)!0.5!(comm.south -| pool)+(0.12,0)$) {shared forecasts};
		\end{tikzpicture}}
		\caption{Architecture of the multi-carrier prosumer community, showing agents, the peer-to-peer pool, the point of common coupling (PCC), the DSO grid interface, and the shared-forecasting platform.}
		\label{fig_arch}
	\end{figure}
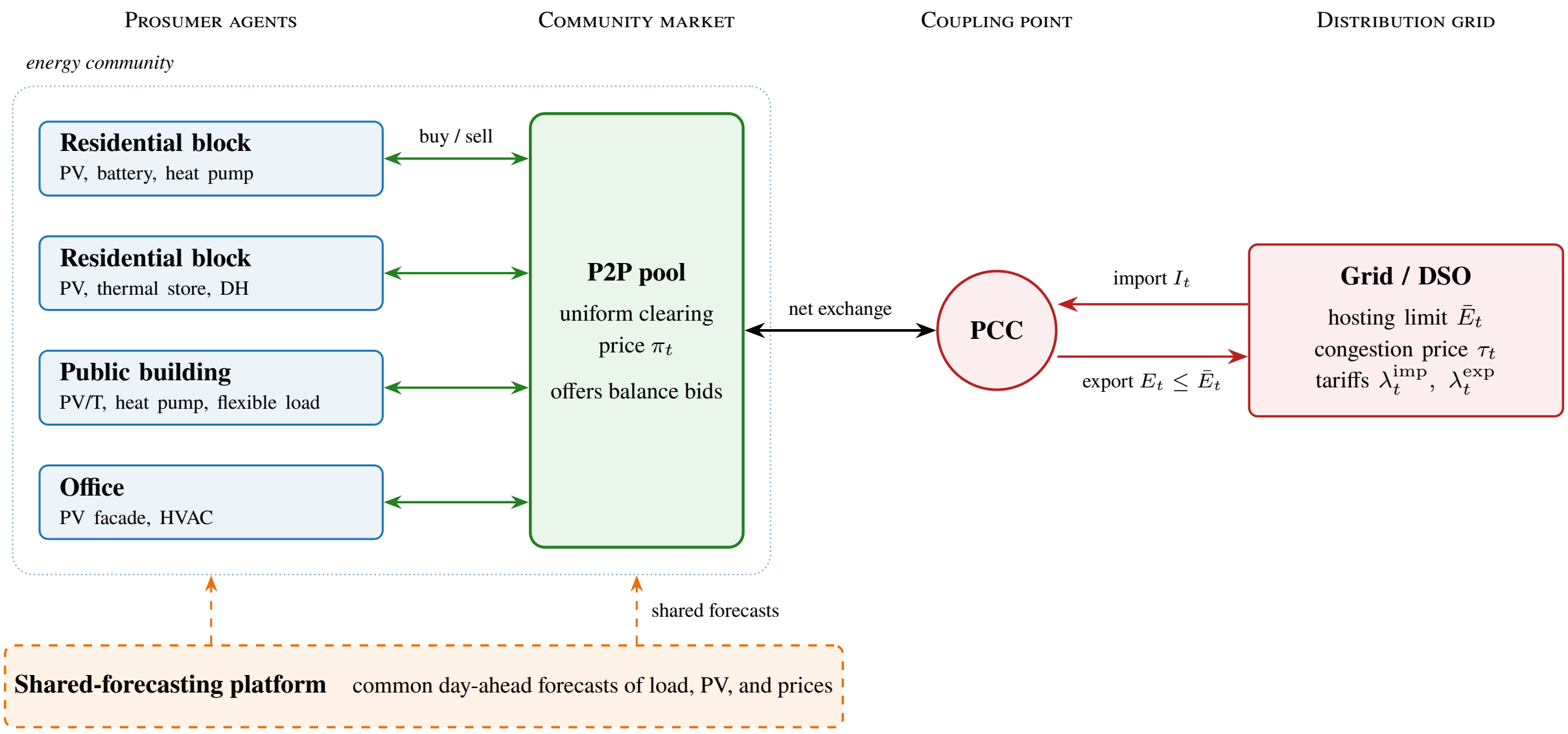

	We consider a positive energy district organized as an energy community of prosumer agents connected to the distribution grid at a single point of common coupling (PCC). The community carries two energy carriers, electricity and heat, coupled by heat pumps and thermal storage. It interacts with a DSO that governs the coupling point through a hosting limit on export and through grid tariffs. Each agent $a\in\mathcal{A}$ owns a subset of technologies, namely rooftop and facade photovoltaics, photovoltaic-thermal collectors, a heat pump, a battery, thermal storage, a district-heating connection where available, and controllable shiftable loads. \Cref{fig_arch} shows the structure of the model. The prosumers exchange surplus with one another through a community pool and reach the grid only through the aggregate net exchange at the coupling point (PCC), which the DSO caps on the export side.
	
	The community can be organized in two ways that we consider side by side. Under central coordination, a community operator co-optimizes every asset with full information and treats internal transfers as costless. Under decentralized coordination, each agent optimizes its own cost, trades surplus in the community pool at a cleared price, and reacts to a grid signal from the DSO. The difference between the two organizations is the coordination gap, and the modeled platform is the set of information and market mechanisms that reduces it. We isolate three such mechanisms. The platform enters this paper only through those three mechanisms. Communication architecture, data governance, cybersecurity, and real-time data exchange are not represented, and no result in this paper speaks to them. \emph{Forecast transparency} lets agents plan against shared forecasts distributed by the platform instead of private, myopic information. An \emph{internal P2P market} allows agents trade surplus with one another before reaching the grid. \emph{Grid-aware pricing} lets the DSO signal congestion, enabling agents internalize the hosting limit, which avoids curtailment. The horizon is discretized into steps $t\in\mathcal{T}=\{1,\dots,T\}$ of length $\Delta t$, assembled from representative weeks per season, and each step carries an annualization weight $w_t$ that scales representative-week quantities to a full year. The primary sets, parameters, and decision variables are collected in \Cref{tab_notation}.
	
	{\setlength{\tabcolsep}{4pt}
		\renewcommand{\arraystretch}{1.08}
		\begin{xltabular}{\textwidth}{@{}l@{\hspace{5pt}}>{\raggedright\arraybackslash\hspace{0pt}}X@{\hspace{12pt}}l@{\hspace{5pt}}>{\raggedright\arraybackslash\hspace{0pt}}X@{}}
			\caption{Notation used throughout the paper.}\label{tab_notation}\\
			\toprule
			\textbf{Symbol} & \textbf{Description} & \textbf{Symbol} & \textbf{Description} \\
			\midrule
			\endfirsthead
			\multicolumn{4}{@{}l}{\itshape\small \tablename~\thetable{} (continued)}\\
			\toprule
			\textbf{Symbol} & \textbf{Description} & \textbf{Symbol} & \textbf{Description} \\
			\midrule
			\endhead
			\bottomrule
			\endlastfoot
			\multicolumn{4}{@{}l}{\textit{Sets, indices, and objects}} \\
			\midrule
			$\mathcal{T},\mathcal{A},\mathcal{S}$ & Time steps of length $\Delta t$, prosumer agents, representative seasons & $t,a,\sigma(t)$ & Index of time step, index of agent, season of step $t$ \\
			$\Xi,\ \mathcal{X}_a$ & Set of all endogenous variables, feasible set of agent $a$ & $x_a$ & Decision vector of agent $a$ ($x_a\in\mathcal{X}_a$) \\
			\midrule
			\multicolumn{4}{@{}l}{\textit{Parameters -- technology and energy-demand}} \\
			\midrule
			$\Delta t,\ w_t$ & Step length, annualization weight of step $t$ [h, --] & $\mathrm{PV}^{\mathrm{av}}_{at}$ & Available photovoltaic power [kW] \\
			$Q^{\mathrm{pvt}}_{at}$ & Photovoltaic-thermal heat [kW] & $D^{\mathrm{fix}}_{at}$ & Fixed (non-shiftable) electrical demand [kW] \\
			$D^{\mathrm{sh}}_{at}$ & Baseline shiftable electrical demand [kW] & $D^{h}_{at}$ & Heat demand (exogenous) [kW] \\
			$T^{\mathrm{amb}}_t$ & Ambient temperature [${}^\circ$C] & $\mathrm{COP}_0,\kappa_{\mathrm{COP}}$ & Heat-pump coefficient-of-performance intercept and slope [--, 1/${}^\circ$C] \\
			$\bar C^B_a,\bar P^B_a$ & Battery energy and power capacity [kWh, kW] & $\underline{\sigma},\bar{\sigma}$ & Battery state-of-charge lower and upper bounds [--] \\
			$\eta_{\mathrm{ch}},\eta_{\mathrm{dis}}$ & Battery charge and discharge efficiency [--] & $\delta_a$ & Battery degradation cost on cycled energy [\euro/kWh] \\
			$\bar C^T_a,\bar P^T_a$ & Thermal-storage energy and power capacity [kWh, kW] & $\eta_T,\kappa_T$ & Thermal-storage efficiency and standing-loss rate [--, 1/h] \\
			$\alpha_a,\bar q^{\mathrm{sh}}_a$ & Shiftable-load flexibility fraction and within-day store capacity [--, kWh] &  &  \\
			$c^{\mathrm{f}},F_a$ & Fabric effective heat capacity and conditioned floor area, fabric sensitivity only [kWh/(m$^2$K), m$^2$] & $\kappa^{\mathrm{f}}_a,\Delta\theta$ & Fabric standing-loss rate and comfort band half-width, fabric sensitivity only [1/h, K] \\
			\midrule
			\multicolumn{4}{@{}l}{\textit{Parameters -- market, grid, and objective}} \\
			\midrule
			$\lambda^{\mathrm{imp}}_t,\lambda^{\mathrm{exp}}_t$ & Grid import and export price [\euro/kWh] & $\tau^{\mathrm{use}}$ & Volumetric use-of-system tariff [\euro/kWh] \\
			$\lambda^{\mathrm{dh}}$ & District-heating energy price [\euro/kWh] & $\mathrm{EF}^{\mathrm{g}}_t,\mathrm{EF}^{\mathrm{dh}}$ & Grid and district-heating emission factors [kg\,CO$_2$/kWh] \\
			$\bar E_t$ & DSO export hosting limit at the coupling point [kW] & $\mu,\varepsilon^{\mathrm{s}},\varepsilon^{\mathrm{d}}$ & Carbon price, spill and heat-dump tie-breaking penalties [\euro/kg\,CO$_2$, \euro/kWh, \euro/kWh] \\
			$\varepsilon^{\mathrm{r}},\epsilon_0$ & Grid- and pool-flow regularizer, matching-rate tolerance [\euro/(kW$^2$h), kW] & $\bar M_a$ & Finite upper bound on the grid and pool flows of agent $a$ [kW] \\
			$\underline{\mathrm{COP}},\overline{\mathrm{COP}}$ & Lower and upper bounds on the heat-pump COP [--] &  &  \\
			\midrule
			\multicolumn{4}{@{}l}{\textit{Decision variables}} \\
			\midrule
			$\mathrm{PV}_{at}$ & Photovoltaic power dispatched [kW] & $\mathrm{ch}^B_{at},\mathrm{dis}^B_{at},e^B_{at}$ & Battery charge, discharge, state of charge [kW, kW, kWh] \\
			$\mathrm{ch}^T_{at},\mathrm{dis}^T_{at},e^T_{at}$ & Thermal-storage charge, discharge, state [kW, kW, kWh] & $P^{\mathrm{he}}_{at},P^{\mathrm{res}}_{at}$ & Heat-pump electrical input, resistive heating input [kW] \\
			$\mathrm{DH}_{at},Q^{\mathrm{d}}_{at}$ & District-heating intake, heat dump [kW] & $d^{\mathrm{sh}}_{at},q^{\mathrm{sh}}_{at}$ & Served shiftable load, shift virtual-store state [kW, kWh] \\
			$\mathrm{imp}_{at},\mathrm{exp}_{at}$ & Agent (metered) grid import and export [kW] & $s_{at},b_{at}$ & Pool sell and buy [kW] \\
			$u^B_{at}$ & Battery non-simultaneity binary [--] & $I_t,E_t,\mathrm{CURT}_t$ & Community net import, net export, curtailed net export [kW] \\
			$e^{\mathrm{f}}_{at},u_{at}$ & Fabric store state and pre-heating charge, fabric sensitivity only [kWh, kW] &  &  \\
			\midrule
			\multicolumn{4}{@{}l}{\textit{Auxiliary quantities, prices, and metrics}} \\
			\midrule
			$\mathrm{COP}_{at}$ & Heat-pump coefficient of performance (from $T^{\mathrm{amb}}_t$) [--] & $L^{\mathrm{spl}}_t$ & Renewable spill (unused PV plus curtailed net export) [kW] \\
			$\pi_t,\tau_t$ & P2P clearing price, congestion price [\euro/kWh] & $\lambda^{\mathrm{pool}}_t,\lambda^{\mathrm{host}}_t$ & Raw duals of pool-balance and net-hosting constraints [\euro/kW] \\
			$\hat\lambda^{\mathrm{pool}}_t,\hat\lambda^{\mathrm{host}}_t$ & Weight-normalized duals, with $\pi_t=-\lambda^{\mathrm{pool}}_t/(w_t\Delta t)$ and $\tau_t=\lambda^{\mathrm{host}}_t/(w_t\Delta t)$ [\euro/kWh] & $D_t,\mathrm{RES}_t$ & Served electrical demand, available renewable generation [kW] \\
			$\mathrm{cost},\mathrm{CO}_2$ & Net cost, carbon emissions [\euro, kg\,CO$_2$] & $\mathrm{SSR},\mathrm{LPI}^{\mathrm{os}},\mathrm{HMR}$ & Self-sufficiency ratio, oversupply loss-of-positivity index, hourly matching rate [--] \\
			$\Gamma_v,\varrho_v$ & Coordination gap and recovered fraction of variant $v$ [\euro, --] &  &  \\
	\end{xltabular}}
	
	\begin{table}[H]
		\centering
		\caption{Provenance of each numbered equation and of the equilibrium result.}
		\label{tab_eqprov}
		\setlength{\tabcolsep}{4pt}
		\begin{tabularx}{\textwidth}{@{}>{\raggedright\arraybackslash}p{1.4cm}>{\raggedright\arraybackslash}p{1.6cm}>{\raggedright\arraybackslash\hspace{0pt}}X@{}}
			\toprule
			Equation & Status & Source and what differs \\
			\midrule
			\eqref{eq_obj} & adapted & Standard prosumer operating cost with a carbon term. The two tie-breaking penalties and the coupling-flow regularizer are this paper's device for a unique allocation. Degradation cost after \citet{xu2018modeling} \\
			\eqref{eq_ebal}, \eqref{eq_hbal} & standard & Energy-hub carrier balances \citep{geidl2007optimal} \\
			\eqref{eq_dump} & this paper & Caps the heat dump by the photovoltaic-thermal input, which prevents electricity being hidden as dumped heat \\
			\eqref{eq_batt}--\eqref{eq_battdis} & standard & Storage dynamics in the energy-hub formulation \citep{geidl2007optimal}, with one-way efficiencies, a state-of-charge window, a power rating, and a non-simultaneity binary \citep{abate2026bidding} \\
			\eqref{eq_tes}--\eqref{eq_tesrate} & standard & Thermal storage with round-trip efficiency and standing loss \citep{sarbu2018comprehensive} \\
			\eqref{eq_shift}--\eqref{eq_pv} & adapted & Shiftable demand after \citet{gils2014assessment}. The modification is the bounded within-day virtual store with closure at every day boundary, which limits both the depth and the duration of a shift \\
			\eqref{eq_pcc} & adapted & Hosting limit at the coupling point, in the spirit of non-firm connection limits \citep{bjarghov2024enhancing}. The net form is this paper's choice, not the gross form \\
			\eqref{eq_pool} & standard & Community pool clearing \citep{sousa2019peer} \\
			\eqref{eq_flowbound} & convention & Finite bounds that make each agent's feasible set compact, needed for \Cref{ass_convex} \\
			\eqref{eq_cyclic}, \eqref{eq_cyclict} & convention & Season-cyclic storage closure, a representative-period modeling convention \\
			\eqref{eq_domain} & standard & Non-negativity and integrality restrictions \\
			\eqref{eq_cop} & standard & Coefficient of performance affine in ambient temperature and clipped \citep{staffell2012review} \\
			\eqref{eq_lpi} & mixed & The self-sufficiency ratio is a standard self-consumption indicator \citep{luthander2015photovoltaic}. The oversupply index is this paper's surplus-side mirror \\
			\eqref{eq_hmr} & this paper & Duration-weighted hourly matching rate, related to the hourly matching literature already cited \\
			\eqref{eq_agent}, \eqref{eq_clearing} & standard & Price-taking agent problem and market clearing \citep{lecadre2020peer} \\
			\eqref{eq_gap} & this paper & Coordination gap and recovered fraction \\
			\eqref{eq_tamb}--\eqref{eq_demeq} & this paper & Deterministic archetype generators for weather, solar, load, heat, and the renewable sizing rule \\
			\bottomrule
		\end{tabularx}
	\end{table}
	
	\subsection{The centralized optimization problem}\label{subsec_optmodel}
	
	The community problem is stated in full below, where its completeness is important. The equilibrium result in Proposition \ref{prop_welfare} depends on which constraints couple the agents and which are private to one agent, and that split is visible only when every constraint is expressed explicitly. The two economic signals of the paper are dual multipliers of two specific constraints, the P2P price of the pool-clearing condition \eqref{eq_pool} and the congestion price of the hosting limit \eqref{eq_pcc}. Each stage of the coordination experiment is then defined by whether the pool and the congestion price are active. That means that the stages are differences between stated programs and not between verbal descriptions. Most of the building blocks are standard, and \Cref{tab_eqprov} gives the provenance of each one. The operator minimizes a monetized objective function, the sum of net grid cost, carbon-priced emissions, and small strictly convex regularizing penalties, subject to the multi-carrier hub constraints of every agent and the grid coupling of the community. Peer-to-peer transfers net out internally under central coordination and drop from the cost. The program is a social-planner problem for the community, although its objective function is the members’ aggregate operating bill under the stated tariffs and settlement.
	
	The objective function \eqref{eq_obj} monetizes the operational indicators. Every term is in \euro{}, the trade-offs between them are read directly in cost, and self-consumption is not a term of the objective function but an outcome of minimizing it. The terms are defined per agent and per time step. The net grid cost is import cost less export revenue plus a volumetric use-of-system tariff $\tau^{\mathrm{use}}$. The battery degradation cost $\delta_a$ prices storage wear on cycled energy. The district-heating energy is charged at price $\lambda^{\mathrm{dh}}$. The carbon cost applies to grid import and district heat at the carbon price $\mu$. Two small tie-breaking penalties, $\varepsilon^{\mathrm{s}}$ on renewable spill $\mathrm{PV}^{\mathrm{av}}_{at}-\mathrm{PV}_{at}$ and $\varepsilon^{\mathrm{d}}$ on heat dump, make the spill and dump split unique. A small regularizer $\varepsilon^{\mathrm{r}}$ on the grid and pool flows makes the relaxed program strictly convex in those flows.  The regularized grid and pool flows are then unique, and offsetting round-trip pool trades are eliminated. Each step is weighted by $w_t\Delta t$, which annualizes representative-week costs. Throughout, the presented \emph{cost} is the \emph{economic objective}, namely the private operating cost (import less export revenue, tariff, degradation, and district-heat energy) plus monetized carbon. It excludes the three small numerical penalties $\varepsilon^{\mathrm{s}},\varepsilon^{\mathrm{d}},\varepsilon^{\mathrm{r}}$, which enter the raw solver objective function only to select a unique allocation. Every coordination gap in this paper is evaluated on this economic cost. With the objective function and the multi-carrier and grid-coupling constraints of every agent, the centralized community operator (the social planner, $\mathrm{SP}$) solves
	
	\begin{subequations}\label{eq_model}
		\begin{align}
			\min_{\Xi}\quad
			& \sum_{t\in\mathcal{T}} w_t\Delta t
			\sum_{a\in\mathcal{A}}\Bigg[
			\underbrace{
				\lambda^{\mathrm{imp}}_t\mathrm{imp}_{at}
				-\lambda^{\mathrm{exp}}_t\mathrm{exp}_{at}
				+\tau^{\mathrm{use}}
				\big(\mathrm{imp}_{at}+\mathrm{exp}_{at}\big)
			}_{\text{Net grid cost}}
			+\underbrace{
				\lambda^{\mathrm{dh}}\mathrm{DH}_{at}
			}_{\text{District heating cost}}
			\nonumber\\
			& \qquad
			+\underbrace{
				\delta_a\big(\mathrm{ch}^B_{at}+\mathrm{dis}^B_{at}\big)
			}_{\text{Battery degradation cost}}
			+\underbrace{
				\mu\big(
				\mathrm{EF}^{\mathrm{g}}_t\mathrm{imp}_{at}
				+\mathrm{EF}^{\mathrm{dh}}\mathrm{DH}_{at}
				\big)
			}_{\text{Carbon cost}}
			\nonumber\\
			& \qquad
			+\underbrace{
				\begin{aligned}
					&\varepsilon^{\mathrm{s}}
					\big(\mathrm{PV}^{\mathrm{av}}_{at}-\mathrm{PV}_{at}\big)
					+\varepsilon^{\mathrm{d}}Q^{\mathrm{d}}_{at}
					\quad +\varepsilon^{\mathrm{r}}\big(
					\mathrm{imp}_{at}^{2}+\mathrm{exp}_{at}^{2}
					+b_{at}^{2}+s_{at}^{2} \big)
				\end{aligned}
			}_{\text{Numerical penalties}} \Bigg]\label{eq_obj}\\\
			\text{s.t.} \nonumber \\
			& \mathrm{PV}_{at}+\mathrm{dis}^B_{at}+\mathrm{imp}_{at}+b_{at}
			= D^{\mathrm{fix}}_{at}+d^{\mathrm{sh}}_{at}+\mathrm{ch}^B_{at}+P^{\mathrm{he}}_{at}+P^{\mathrm{res}}_{at}+\mathrm{exp}_{at}+s_{at}, \label{eq_ebal}\\
			& \mathrm{COP}_{at}P^{\mathrm{he}}_{at}+P^{\mathrm{res}}_{at}+Q^{\mathrm{pvt}}_{at}+\mathrm{dis}^T_{at}+\mathrm{DH}_{at}
			= D^{h}_{at}+\mathrm{ch}^T_{at}+Q^{\mathrm{d}}_{at}, \label{eq_hbal}\\
			& 0\le Q^{\mathrm{d}}_{at}\le Q^{\mathrm{pvt}}_{at}, \label{eq_dump}\\
			& e^B_{a,t+1}=e^B_{at}+\eta_{\mathrm{ch}}\mathrm{ch}^B_{at}\Delta t-\tfrac{1}{\eta_{\mathrm{dis}}}\mathrm{dis}^B_{at}\Delta t, \label{eq_batt}\\
			& \underline{\sigma}\bar C^B_a\le e^B_{at}\le\bar{\sigma}\bar C^B_a, \label{eq_battsoc}\\
			& \mathrm{ch}^B_{at}\le\bar P^B_a u^B_{at}, \label{eq_battch}\\
			& \mathrm{dis}^B_{at}\le\bar P^B_a(1-u^B_{at}), \label{eq_battdis}\\
			& e^T_{a,t+1}=(1-\kappa_T\Delta t)e^T_{at}+\eta_T\mathrm{ch}^T_{at}\Delta t-\tfrac{1}{\eta_T}\mathrm{dis}^T_{at}\Delta t, \label{eq_tes}\\
			& 0\le e^T_{at}\le\bar C^T_a, \label{eq_tesbound}\\
			& 0\le\mathrm{ch}^T_{at},\mathrm{dis}^T_{at}\le\bar P^T_a, \label{eq_tesrate}\\
			& q^{\mathrm{sh}}_{a,t+1}=q^{\mathrm{sh}}_{at}+(D^{\mathrm{sh}}_{at}-d^{\mathrm{sh}}_{at})\Delta t, \label{eq_shift}\\
			& -\bar q^{\mathrm{sh}}_a\le q^{\mathrm{sh}}_{at}\le\bar q^{\mathrm{sh}}_a, \label{eq_shiftbound}\\
			& q^{\mathrm{sh}}_{a,24j}=0, \label{eq_shiftclose}\\
			& |d^{\mathrm{sh}}_{at}-D^{\mathrm{sh}}_{at}|\le\alpha_a(D^{\mathrm{fix}}_{at}+D^{\mathrm{sh}}_{at}), \label{eq_dev}\\
			& 0\le\mathrm{PV}_{at}\le\mathrm{PV}^{\mathrm{av}}_{at}, \label{eq_pv}\\
			& \textstyle\sum_{a\in\mathcal{A}}\big(\mathrm{exp}_{at}-\mathrm{imp}_{at}\big)\le\bar E_t, \label{eq_pcc}\\
			& \textstyle\sum_{a\in\mathcal{A}} s_{at}=\sum_{a\in\mathcal{A}} b_{at}, \label{eq_pool}\\
			& 0\le\mathrm{imp}_{at},\,\mathrm{exp}_{at},\,b_{at},\,s_{at}\le\bar M_a, \label{eq_flowbound}\\
			& e^B_{aS\ell}=e^B_{a0},\qquad \ell=1,\dots,|\mathcal{S}|, \label{eq_cyclic}\\
			& e^T_{aS\ell}=e^T_{a0},\qquad \ell=1,\dots,|\mathcal{S}|, \label{eq_cyclict}\\
			& \mathrm{PV}_{at}, \mathrm{ch}^B_{at}, \mathrm{dis}^B_{at}, e^B_{at}, \mathrm{ch}^T_{at}, \mathrm{dis}^T_{at}, e^T_{at} \ge 0, \nonumber\\
			& P^{\mathrm{he}}_{at}, P^{\mathrm{res}}_{at}, \mathrm{DH}_{at}, d^{\mathrm{sh}}_{at}, \mathrm{imp}_{at}, \mathrm{exp}_{at}, s_{at}, b_{at} \ge 0, \quad u^B_{at}\in\{0,1\}. \label{eq_domain}
		\end{align}
	\end{subequations}
	
	where
	$\Xi=\{\mathrm{PV}_{at},\mathrm{ch}^B_{at},\mathrm{dis}^B_{at},e^B_{at},\mathrm{ch}^T_{at},\mathrm{dis}^T_{at},e^T_{at},P^{\mathrm{he}}_{at},P^{\mathrm{res}}_{at},\mathrm{DH}_{at},Q^{\mathrm{d}}_{at},d^{\mathrm{sh}}_{at},q^{\mathrm{sh}}_{at},\mathrm{imp}_{at},\mathrm{exp}_{at},s_{at},b_{at},u^B_{at}\}$
	
	is the set of decision variables. The only integrality is the battery non-simultaneity binary $u^B_{at}$ in \eqref{eq_battch} and \eqref{eq_battdis}, and the only nonlinearity is the small convex regularizer $\varepsilon^{\mathrm{r}}$ on the grid and pool flows. Therefore, the program is an mixed integer quadratic problem (MIQP). Its continuous relaxation, denoted $\mathrm{SP}_{\mathrm{rel}}$, relaxes the binary to $[0,1]$. It is then a convex quadratic program, strictly convex in the grid and pool flows, whose regularized grid and pool flows are unique and whose optimal value is a lower bound on the mixed-integer objective function. A bound on annual net import traces the trade-off between cost and self-sufficiency. The heat-pump coefficient of performance in \eqref{eq_hbal} is a temperature-dependent parameter,
	\begin{equation}\label{eq_cop}
		\mathrm{COP}_{at}=\operatorname{clip}\!\big(\mathrm{COP}_0+\kappa_{\mathrm{COP}}(T^{\mathrm{amb}}_t-7),\ \underline{\mathrm{COP}},\ \overline{\mathrm{COP}}\big),\qquad \operatorname{clip}(x,\ell,u)=\min\{\max\{x,\ell\},u\},
	\end{equation}
	which depends only on the exogenous ambient temperature and enters the heat balance as a fixed coefficient, keeping the program linear in the decision variables. The constraints hold for every agent $a\in\mathcal{A}$ and step $t\in\mathcal{T}$ unless a quantifier states otherwise. Constraint \eqref{eq_ebal} is the agent electricity balance, in which photovoltaic dispatch, battery discharge, grid import, and pool purchases meet fixed and shifted load, battery charge, heat-pump and resistive electricity, grid export, and pool sales. Constraint \eqref{eq_hbal} is the heat balance, met by the heat pump at the temperature-dependent coefficient of performance \eqref{eq_cop}, resistive heat, photovoltaic-thermal heat, thermal-storage discharge, and district heating, with the heat demand $D^h_{at}$ exogenous. Renewable curtailment is thereby represented explicitly, and any available photovoltaic power that is not dispatched, stored, consumed, or exported appears as spill $\mathrm{PV}^{\mathrm{av}}_{at}-\mathrm{PV}_{at}$.
	Constraints \eqref{eq_batt} to \eqref{eq_battdis} govern the battery through one-way charge and discharge efficiencies, a state-of-charge window, a power rating, and the binary $u^B_{at}$ that forbids simultaneous charge and discharge, retained for generality. Simultaneous charge and discharge is not observed in any computed solution, being discouraged by efficiency losses and the degradation cost, and is checked ex post. Constraints \eqref{eq_tes} to \eqref{eq_tesrate} govern the thermal storage, with a round-trip efficiency $\eta_T$ and an hourly standing loss $\kappa_T$. Constraints \eqref{eq_shift} to \eqref{eq_shiftclose} model the shiftable load as a bounded virtual store $q^{\mathrm{sh}}_{at}$ that accumulates the difference between baseline and served load, is bounded by $\bar q^{\mathrm{sh}}_a$, and is closed at every day boundary $24j$, which shifts load only within the day and within a bounded duration. The store may go negative to its bound, permitting shiftable load to be served ahead of baseline, an intended pre-serving of the targeted flexible loads. Building thermal mass is not represented in the base model. A sensitivity switches on a bounded within-day fabric store for each heat-pump agent, mirroring the shiftable-load store of \eqref{eq_shift}, in which pre-heating raises a fabric state $e^{\mathrm{f}}_{at}$ that decays at the inverse building time constant, $e^{\mathrm{f}}_{a,t+1}=(1-\kappa^{\mathrm{f}}_a\Delta t)e^{\mathrm{f}}_{at}+u_{at}\Delta t$ with $0\le e^{\mathrm{f}}_{at}\le c^{\mathrm{f}}F_a\Delta\theta$, closure at every day boundary, and the heat-pump thermal rating holding on served heat plus charge, $D^{h}_{at}+u_{at}\le q^{\mathrm{des}}_a$. The switch is off in every other result. Constraints \eqref{eq_dev} and \eqref{eq_pv} limit the hourly load deviation by the flexibility fraction $\alpha_a$ and cap photovoltaic dispatch at the available power $\mathrm{PV}^{\mathrm{av}}_{at}$.
	Constraint \eqref{eq_pcc} is the net-export hosting limit at the point of common coupling. The prosumers share this coupling point, although they are separately metered. Each settles its own import and export with the retailer at the retail and feed-in prices, which is the reason the objective function sums agent-level flows. The transformer, however, carries only the \emph{net} community injection. The constraint caps net export $\sum_a(\mathrm{exp}_{at}-\mathrm{imp}_{at})\le\bar E_t$, not gross export. Internal exchange between prosumers, whether physical through the low-voltage feeder or financial through the pool \eqref{eq_pool}, does not cross the transformer. This single cap is a reduced representation of the interface with the distribution system operator. It does not represent feeder voltage, network losses, transformer thermal loading, or any limit on the import side. With retail above feed-in no agent both imports and exports in the same step. Gross and net coincide when the community is coordinated, and they diverge only when uncoordinated prosumers meet the grid separately, which is the value the pool captures.\footnote{Retail exceeds feed-in in every hour, including at the diurnal price extremes of \Cref{tab_tech}. The retail-minus-feed-in spread $\lambda^{\mathrm{imp}}_t+\tau^{\mathrm{use}}-\lambda^{\mathrm{exp}}_t=(1-\lambda^{\mathrm{exp}}/\lambda^{\mathrm{imp}})\lambda^{\mathrm{imp}}_t+\tau^{\mathrm{use}}$ is strictly positive for every zone, since the export ratio is below one, the tariff is nonnegative, and the import price stays positive at its diurnal trough.} Constraint \eqref{eq_pool} clears the community pool internally at every step, and aggregate pool sales equal aggregate pool purchases. Constraints \eqref{eq_cyclic} and \eqref{eq_cyclict} make each representative season storage-neutral by pinning the storage level at every season boundary $S\ell$ to the common start level, which prevents non-physical inter-seasonal state transfer. Constraint \eqref{eq_domain} states the non-negativity and integrality restrictions on the decision variables. 
	
	\subsection{Performance indicators}\label{subsec_lpi}
	
	Three operational indicators summarize district performance. The net cost and the carbon emissions are the monetary and carbon parts of the objective function \eqref{eq_obj}. Self-consumption is measured on the community net grid exchange. Let the net community draw at step $t$ be $n_t=\sum_a(\mathrm{imp}_{at}-\mathrm{exp}_{at})$, the net import $I_t=\max(n_t,0)$, and the net export $E_t=\max(-n_t,0)$. The spill is $L^{\mathrm{spl}}_t=\sum_a(\mathrm{PV}^{\mathrm{av}}_{at}-\mathrm{PV}_{at})+\mathrm{CURT}_t$. In this expression $\mathrm{PV}_{at}$ is the photovoltaic dispatch chosen in the optimization, before any recourse. The difference $\mathrm{PV}^{\mathrm{av}}_{at}-\mathrm{PV}_{at}$ is the optimization spill, while $\mathrm{CURT}_t$ is the \emph{additional} net export curtailed by the DSO recourse. The two are disjoint, and the available renewable energy obeys the identity $\sum_a\mathrm{PV}^{\mathrm{av}}_{at}=(\text{self-consumed}) + (\text{effective export}) + L^{\mathrm{spl}}_t$, with no double counting. The oversupply index combines a remunerated export component and a curtailed or spilled component with opposite welfare signs, and we reserve the terms inefficiency and burden for the latter, reading remunerated export as exposed to but not necessarily costly for the grid. For this accounting we attribute net export to renewable generation, exact whenever net export does not exceed the contemporaneous renewable surplus. The model does not track the origin of stored electricity, which a formal provenance claim would require. Let $D_t$ be the served electrical demand and $\mathrm{RES}_t$ the available renewable generation.
	We present two energy ratios at hourly resolution, the self-sufficiency ratio ($\mathrm{SSR}$), the share of served demand met without net grid import, and its surplus-side mirror, the oversupply loss-of-positivity index ($\mathrm{LPI}^{\mathrm{os}}$), the share of available renewable generation exported or curtailed instead of self-consumed,
	
	\begin{equation}\label{eq_lpi}
		\mathrm{SSR}=1-\frac{\sum_t I_t w_t\Delta t}{\sum_t D_t w_t\Delta t},\qquad
		\mathrm{LPI}^{\mathrm{os}}=\frac{\sum_t (E_t+L^{\mathrm{spl}}_t) w_t\Delta t}{\sum_t\mathrm{RES}_t w_t\Delta t}.
	\end{equation}
	
	The step weight $w_t\Delta t$ stays inside the sums. The weight $w_t$ is the seasonal annualization weight and differs across the four representative weeks. Each ratio is an annualized energy-weighted average over the four representative weeks. Since these are energy-weighted, we also present the hourly matching rate ($\mathrm{HMR}$), the duration-weighted share of \emph{hours} met without net import,
	
	\begin{equation}\label{eq_hmr}
		\mathrm{HMR}=\frac{\sum_t \mathbf{1}[I_t\le\epsilon_0]\, w_t}{\sum_t w_t},
	\end{equation}
	
	where $\epsilon_0$ is a small numerical tolerance. $\mathrm{SSR}$ and $\mathrm{LPI}^{\mathrm{os}}$ are energy-weighted, the first by served demand and the second by available renewable generation. $\mathrm{HMR}$ counts hours and is a genuine hour-share indicator. The three are not comparable as shares, and an hour of small deficit counts the same as an hour of large deficit in $\mathrm{HMR}$ alone.
	The residual hourly imbalance is the pair $(1-\mathrm{SSR},\mathrm{LPI}^{\mathrm{os}})$. Annual positivity requires $\sum_t\mathrm{RES}_t w_t\Delta t\ge\mathcal{D}$, with $\mathcal{D}$ the annual electricity-equivalent demand of \eqref{eq_demeq}, which carries the same step weights and converts heat at the average heat-pump coefficient of performance. A district can satisfy it and still show a large deficit $1-\mathrm{SSR}$ in winter and night hours together with a large oversupply $\mathrm{LPI}^{\mathrm{os}}$ at summer middays. Reducing one component without inflating the other is the coordination problem that this paper addresses.
	
	\subsection{The decentralized P2P equilibrium}\label{subsec_equilibrium}
	
	Under decentralized coordination each agent solves its own cost minimization problem, taking the P2P price $\pi_t$ and the DSO congestion price $\tau_t$ as given,
	
	\begin{equation}\label{eq_agent}
		\begin{split}
			\min_{x_a\in\mathcal{X}_a}\ \sum_t w_t\Delta t\Big[ & \lambda^{\mathrm{imp}}_t\mathrm{imp}_{at}-\lambda^{\mathrm{exp}}_t\mathrm{exp}_{at}+\tau^{\mathrm{use}}(\mathrm{imp}_{at}+\mathrm{exp}_{at})
			+\pi_t(b_{at}-s_{at})+\tau_t(\mathrm{exp}_{at}-\mathrm{imp}_{at}) \\
			& +\delta_a(\mathrm{ch}^B_{at}+\mathrm{dis}^B_{at})+\lambda^{\mathrm{dh}}\mathrm{DH}_{at}
			+\mu\big(\mathrm{EF}^{\mathrm{g}}_t\mathrm{imp}_{at}+\mathrm{EF}^{\mathrm{dh}}\mathrm{DH}_{at}\big) \\
			& +\varepsilon^{\mathrm{s}}(\mathrm{PV}^{\mathrm{av}}_{at}-\mathrm{PV}_{at})+\varepsilon^{\mathrm{d}}Q^{\mathrm{d}}_{at}
			+\varepsilon^{\mathrm{r}}\big(\mathrm{imp}_{at}^2+\mathrm{exp}_{at}^2+b_{at}^2+s_{at}^2\big)\Big],
		\end{split}
	\end{equation}
	
	where $\mathcal{X}_a$ is the agent's own feasible set, namely the agent-private rows of \eqref{eq_model}. Every term matches the corresponding per-agent term of the planner objective function \eqref{eq_obj}. The agent problem \eqref{eq_agent} is exactly the agent block of the partial Lagrangian obtained by dualizing the two coupling constraints, the pool balance \eqref{eq_pool} and the net-export hosting limit \eqref{eq_pcc}. The agent pays the P2P price $\pi_t$ on its net pool position $b_{at}-s_{at}$ and the congestion price $\tau_t$ on its \emph{net} grid export $\mathrm{exp}_{at}-\mathrm{imp}_{at}$, consistent with the net form of the hosting limit, and it takes both prices as parametric. The market clears when the pool balances and the aggregate net export respects the hosting limit,
	
	\begin{subequations}\label{eq_clearing}
		\begin{align}
			& \sum_a s_{at}=\sum_a b_{at}\ \Rightarrow\ \pi_t, \label{eq_clearpool}\\
			& \sum_a(\mathrm{exp}_{at}-\mathrm{imp}_{at})\le\bar E_t,\ \tau_t\ge 0,\ \text{complementary}\ \Rightarrow\ \tau_t. \label{eq_clearhost}
		\end{align}
	\end{subequations}
	
	A tuple $(x^*,\pi^*,\tau^*)$ in which each agent solves \eqref{eq_agent} and the clearing conditions \eqref{eq_clearing} hold is a competitive equilibrium of the community market. We make two standard assumptions.
	
	\begin{ass}[Convex prosumers]\label{ass_convex}
		The battery non-simultaneity binaries are relaxed to the unit interval, which makes each feasible set $\mathcal{X}_a$ a convex polytope. The device flows are bounded by the finite installed capacity and demand. The grid and pool flows $(\mathrm{imp}_{at},\mathrm{exp}_{at},b_{at},s_{at})$ are given explicit finite upper bounds, set to a generous multiple of the peak feasible flow, which makes $\mathcal{X}_a$ compact. A small quadratic regularizer $\varepsilon^{\mathrm{r}}$ (with units of \euro/(kW$^2$ h) for hourly power variables) is added on the grid and pool flows. It penalizes gross trading, which eliminates any offsetting round-trip in $(b_{at},s_{at})$ and makes the regularized grid and pool flows unique. The storage, heating, and other internal device dispatch that support them may remain nonunique.
	\end{ass}
	
	\begin{ass}[Price-taking]\label{ass_pricetaking}
		Each agent treats the P2P price $\pi_t$ and the congestion price $\tau_t$ as given and does not act strategically to influence them.
	\end{ass}
	
	The relaxation in \Cref{ass_convex} is mild for prosumer operation, since a domestic controller does not solve a mixed-integer program, and simultaneous charge and discharge is discouraged by the round-trip efficiencies and the degradation cost and is zero in every computed solution. Under these assumptions the equilibrium allocation solves $\mathrm{SP}_{\mathrm{rel}}$, and the economic prices are the coupling-constraint multipliers of the community problem.
	
	\begin{prop}[Competitive implementation of the regularized community optimum]\label{prop_welfare}
		Under \Cref{ass_convex,ass_pricetaking}, a competitive equilibrium of \eqref{eq_agent} and \eqref{eq_clearing} exists whenever $\mathrm{SP}_{\mathrm{rel}}$ is feasible, and its allocation solves $\mathrm{SP}_{\mathrm{rel}}$. As all constraints are polyhedral, strong duality holds without a constraint qualification. Every equilibrium has the same grid and pool flows $(\mathrm{imp}^*,\mathrm{exp}^*,b^*,s^*)$, which are the unique optimal flows of the regularized problem, and the same optimal cost. The internal device dispatch need not be unique. Supporting prices exist and equal the \emph{weight-normalized} Lagrange multipliers of the pool-balance constraint \eqref{eq_pool} and the net-export hosting limit \eqref{eq_pcc}. Since the objective function carries the annualization weight $w_t\Delta t$ while the coupling constraints do not, the economic prices in \euro{} per kilowatt hour are $\pi^*_t=\hat\lambda^{\mathrm{pool}}_t\equiv-\lambda^{\mathrm{pool}}_t/(w_t\Delta t)$ and $\tau^*_t=\hat\lambda^{\mathrm{host}}_t\equiv\lambda^{\mathrm{host}}_t/(w_t\Delta t)$, where $\lambda^{\mathrm{pool}}_t$ and $\lambda^{\mathrm{host}}_t$ are the raw multipliers and the sign on the pool price follows the orientation of \eqref{eq_pool}. The prices need not be unique.
	\end{prop}
	
	\begin{prof}
		Under \Cref{ass_convex} each feasible set $\mathcal{X}_a$ is a compact convex polytope and each agent objective function is convex, strictly so in the grid and pool flows through the regularizer $\varepsilon^{\mathrm{r}}$. Therefore, $\mathrm{SP}_{\mathrm{rel}}$ is a convex quadratic program. The resistive backup and grid import make every agent feasible, and an optimum exists with a unique regularized coupling-flow allocation. The only constraints that couple agents are the pool balance \eqref{eq_pool} and the net-export hosting limit \eqref{eq_pcc}, and every other constraint is agent-private. Dualizing these two families with multipliers $\lambda^{\mathrm{pool}}_t\in\R$ and $\lambda^{\mathrm{host}}_t\ge 0$ gives the partial Lagrangian

		$$\mathcal{L}=\sum_a f_a(x_a)+\sum_t\lambda^{\mathrm{pool}}_t\sum_a(s_{at}-b_{at})+\sum_t\lambda^{\mathrm{host}}_t\Big(\sum_a(\mathrm{exp}_{at}-\mathrm{imp}_{at})-\bar E_t\Big),$$
		
		where $f_a$ is agent $a$’s objective function, the per-agent summand of \eqref{eq_obj}, and $\mathcal{L}$ separates across agents. The objective function $f_a$ carries the weight $w_t\Delta t$ in every step while the coupling terms do not. Writing $\lambda^{\mathrm{pool}}_t=-w_t\Delta t\,\hat\lambda^{\mathrm{pool}}_t$ and $\lambda^{\mathrm{host}}_t=w_t\Delta t\,\hat\lambda^{\mathrm{host}}_t$ places the coupling terms under the same weight, and agent $a$’s part of $\mathcal{L}$ becomes the agent problem \eqref{eq_agent} with $\pi_t=\hat\lambda^{\mathrm{pool}}_t$ charged on the net bought position $b_{at}-s_{at}$ and $\tau_t=\hat\lambda^{\mathrm{host}}_t$ charged on the net export $\mathrm{exp}_{at}-\mathrm{imp}_{at}$. The remaining carbon, spill, dump, and regularizer terms are agent-private and identical in \eqref{eq_agent} and \eqref{eq_obj}. As a feasible and bounded convex quadratic program with polyhedral constraints, $\mathrm{SP}_{\mathrm{rel}}$ satisfies strong duality without a further constraint qualification, with optimal primal-dual pair $(x^*,\lambda^{\mathrm{pool}},\lambda^{\mathrm{host}})$. Its KKT system, each agent’s private stationarity and feasibility together with the pool-balance equality and the hosting complementarity $\lambda^{\mathrm{host}}_t\big(\bar E_t-\sum_a(\mathrm{exp}_{at}-\mathrm{imp}_{at})\big)=0$ with $\lambda^{\mathrm{host}}_t\ge0$, is exactly the equilibrium system \eqref{eq_agent} and \eqref{eq_clearing} at $(\pi^*,\tau^*)=(\hat\lambda^{\mathrm{pool}},\hat\lambda^{\mathrm{host}})$, and $x^*$ with these prices is a competitive equilibrium. Conversely, any equilibrium satisfies the same system and, by strict convexity in the grid and pool flows, shares the unique coupling-flow allocation $(\mathrm{imp}^*,\mathrm{exp}^*,b^*,s^*)$ and the optimal cost. The supporting prices exist by strong duality but need not be unique.
	\end{prof}
	
	The equivalence is exact for the regularized objective function. Therefore, the competitive equilibrium implements the regularized community optimum. The economic cost we present removes the numerical spill, dump, and regularizer penalties. It is the associated unregularized cost, which differs from the regularized optimum only by those tie-breaking terms. As $\varepsilon^{\mathrm{r}}\to 0$, the regularized grid and pool flows converge to the grid-and-pool-flow vector that minimizes the representative-week-weighted Euclidean norm $\sum_t w_t(\mathrm{imp}_{at}^2+\mathrm{exp}_{at}^2+\mathrm{buy}_{at}^2+\mathrm{sell}_{at}^2)$ among the optimal solutions of the problem without the flow regularizer, the quadratic penalty carrying the same weights $w_t$ as the objective function. Only the flow regularizer vanishes in that limit. The spill and dump tie-breaking penalties are fixed constants and are held at their values throughout, so the limiting problem is not the penalty-free one. The regularizer does not select a least-norm solution for the internal storage, heating, and demand-response variables, which may remain nonunique. Under suitable regularity, accumulation points of bounded selections of the supporting prices lie in the unregularized dual solution set. The presented prices are shadow values of the physical constraints up to the vanishing tie-breaking term. \Cref{prop_welfare} then has two consequences that structure the analysis.
	
	\begin{remark}[Separation of the two sides of the gap]\label{rem_separation}
		\Cref{prop_welfare} concerns the organizational cost gap under the convex perfect-information stages. It does not bound physical self-sufficiency. That is limited by installed storage and flexibility, not by organization. The perfect-information market-and-grid-price benchmark closes the cost gap, whereas self-sufficiency remains bounded by storage. The shared-forecast stage is presented separately and does not, by itself, implement the perfect-information optimum. Convexity gives the exact equivalence between the equilibrium and $\mathrm{SP}_{\mathrm{rel}}$. Without it the equilibrium reaches only the relaxation, and the difference between $\mathrm{SP}$ and $\mathrm{SP}_{\mathrm{rel}}$ cannot be supported by uniform linear prices. Price taking lets $\pi_t$ and $\tau_t$ enter each agent’s problem as parameters. Without it a large exporter would act on its own effect on the congestion price or the pool price, and the equilibrium would move away from $\mathrm{SP}_{\mathrm{rel}}$.
	\end{remark}
	
	First, the equilibrium can be computed as the convex community problem, and the economic prices are recovered from the coupling-constraint multipliers, which avoids relying on an iterative price search for any headline result. Second, the coordination gap between a decentralized community and the planner comes from a small number of identifiable frictions. It does not come from the market mechanism itself. When agents plan on private and myopic information, when a flat tariff fails to price the hosting limit, and when surplus cannot be traded internally, the decentralized outcome departs from $\mathrm{SP}_{\mathrm{rel}}$, and each departure is attributable to one missing mechanism. An integrality gap from $\mathrm{SP}_{\mathrm{rel}}$ to the full mixed-integer planner $\mathrm{SP}$ remains from the binary non-simultaneity decisions of the batteries. By \Cref{prop_welfare} this gap is not generally implementable by uniform linear marginal prices without side payments or an alternative settlement.
	
	\subsection{The mechanism waterfall, solution method, and computation}\label{subsec_waterfall}
	
	We evaluate coordination through five decentralized stages taken in a fixed order, which we call the waterfall, and consider as two experiments. The first experiment is forecast-informed and runs from $\mathrm{DE0}$ to $\mathrm{M1}$, the only stage where agents plan under forecast error. The second is a perfect-information decomposition that takes $\mathrm{IND}$ as its baseline and adds the pool and then the grid price. The distance between the two, $\mathrm{M1}$ to $\mathrm{IND}$, is not a mechanism contribution. It is the unattributed residual that separates a forecast-informed stage from a perfect-information one. Each experiment is measured against its own baseline. The forecast-informed step is a share of the $\mathrm{DE0}$-to-$\mathrm{SP}$ gap, and the pool and grid-price steps are shares of the $\mathrm{IND}$-to-$\mathrm{SP}$ distance. Shares taken against different baselines are not additive. Since sequential marginal contributions can depend on their order, the waterfall is an interpretable attribution path. It is not a unique decomposition of intrinsic mechanism values.
	
	\begin{definition}[Mechanism waterfall]\label{def_waterfall}
		
		The five decentralized variants are short labels for the stages of the platform waterfall. They are the decentralized no-platform baseline $\mathrm{DE0}$ (zero mechanisms), the three platform mechanisms $\mathrm{M1}$, $\mathrm{M2}$, and $\mathrm{M3}$ (shared forecasting, P2P market clearing, and grid-aware pricing, added in that order), and the independent perfect-information control $\mathrm{IND}$, the baseline of the second experiment. They are defined as follows.
		\begin{itemize}
			\item[$\mathrm{DE0}$] \emph{no platform}. Each agent optimizes alone over a daily rolling horizon under a flat tariff and no pool, planning on its own \emph{private} day-ahead forecast and realizing on the true data, with the hosting limit enforced by the DSO recourse. The committed intertemporal decision is the battery schedule, while thermal storage, heat-pump and grid dispatch, and the within-day shiftable load are re-optimized against the realized day. The realization is a forecast-informed battery commitment under perfect intraday recourse, not a real-time receding-horizon controller. Battery and thermal storage are reset to a reference state at each representative-week boundary, matching the season-cyclic closure of the planner and perfect-information variants. \Cref{alg_rolling} in \ref{app_formulation} states the full procedure.
			\item[$\mathrm{M1}$] \emph{transparency (value of shared information)}. Identical to $\mathrm{DE0}$ except that agents plan on a \emph{shared}, more accurate forecast distributed by the platform, over the same daily rolling horizon. Since the platform both raises accuracy and makes the forecast common across agents, the $\mathrm{DE0}\!\to\!\mathrm{M1}$ increment is the combined value of higher accuracy and shared consistency, not of accuracy alone.\footnote{The presented $\mathrm{M1}$ value isolates forecast-informed battery commitment while other operational decisions are treated as recourse, and it should not be interpreted as the value of forecasting every multi-carrier decision.}
			\item[$\mathrm{IND}$] \emph{full foresight (baseline of the second experiment)}. Each agent optimizes alone over the full horizon under \emph{perfect} information, with no pool, hosting by recourse, and the \emph{same season-cyclic storage} as $\mathrm{M2}$ and $\mathrm{M3}$. It is a diagnostic control, not a deployable mechanism. It absorbs the value of perfect foresight and the full horizon, and the pool step that follows then measures the value of trading alone.
			\item[$\mathrm{M2}$] \emph{P2P market (perfect-information stage)}. Agents trade surplus in the community pool at the cleared price $\pi_t$ under perfect information and a flat grid tariff, while the net-export hosting limit is enforced through recourse.
			\item[$\mathrm{M3}$] \emph{grid-aware pricing (perfect-information stage)}. The DSO congestion price $\tau_t$ internalizes the net-export hosting limit. By \Cref{prop_welfare}, the outcome equals $\mathrm{SP}_{\mathrm{rel}}$.
		\end{itemize}
	\end{definition}
	
	The first experiment yields one value, the shared-forecast contribution $\mathrm{DE0}\!\to\!\mathrm{M1}$, measured against the $\mathrm{DE0}$-to-$\mathrm{SP}$ gap. The second yields two, P2P organization $\mathrm{IND}\!\to\!\mathrm{M2}$ and grid-aware pricing $\mathrm{M2}\!\to\!\mathrm{M3}$, each measured against the $\mathrm{IND}$-to-$\mathrm{SP}$ distance that its own baseline defines. Since $\mathrm{IND}$ and $\mathrm{M2}$ share perfect information and season-cyclic storage, $\mathrm{IND}\!\to\!\mathrm{M2}$ changes only the organization. By contrast, $\mathrm{M1}\!\to\!\mathrm{IND}$ changes forecast accuracy, horizon length, and the storage boundary convention. It is a diagnostic residual, not a causal estimate of foresight. The coordination gap and the recovered fraction are
	\begin{equation}\label{eq_gap}
		\Gamma_v=\mathrm{cost}(v)-\mathrm{cost}(\mathrm{SP}),\qquad
		\varrho_v=\frac{\Gamma_{\mathrm{DE0}}-\Gamma_v}{\Gamma_{\mathrm{DE0}}},
	\end{equation}
	and the recovered fraction at $\mathrm{M3}$ is the recovery of the perfect-information stages, not the shared-forecast benefit at $\mathrm{M1}$. Inside the second experiment the same fraction is taken against its own baseline, $\varrho^{\mathrm{IND}}_v=(\Gamma_{\mathrm{IND}}-\Gamma_v)/\Gamma_{\mathrm{IND}}$ for $v\in\{\mathrm{M2},\mathrm{M3}\}$. Both normalizations are given, the first against $\mathrm{DE0}$ and the second against $\mathrm{IND}$, which are the baselines of the two experiments.
	
	\Cref{alg_waterfall} is the master procedure behind every headline result. The shared-forecast step changes the quality and cross-agent consistency of the shared forecast while holding the rolling horizon, physical assets, tariffs, and recourse structure fixed. The second experiment then adopts full-horizon perfect information and the seasonal storage convention used by the joint variants. This ensures that the following P2P comparison isolates the organizational change. Grid-aware pricing internalizes the net-export cap. In variants without that price, the DSO recourse in \Cref{subsec_calibration} reduces dispatched photovoltaic power and export by the same amount, preserving feasibility. The economic equilibrium prices are the coupling-constraint multipliers from \Cref{prop_welfare}.
	\begin{algorithm}[H]
		\caption{Controlled coordination-gap waterfall for one district configuration}\label{alg_waterfall}
		\begin{algorithmic}[1]
			\Require calibrated profiles, technology and market parameters, hosting limit $\bar E_t$
			\Ensure KPIs and prices for $\{\mathrm{SP},\mathrm{DE0},\mathrm{M1},\mathrm{IND},\mathrm{M2},\mathrm{M3}\}$ and the gap decomposition
			\State Solve the planner \eqref{eq_model} and its relaxation $\mathrm{SP}_{\mathrm{rel}}$
			\State $\mathrm{DE0}$: per agent, roll a daily horizon; plan \eqref{eq_agent} on a \emph{private} forecast, realize on true data; aggregate; enforce $\bar E_t$ by the hosting recourse
			\State $\mathrm{M1}$: as $\mathrm{DE0}$ with a \emph{shared} forecast at the same horizon
			\State $\mathrm{IND}$: per agent, solve \eqref{eq_agent} once over the full horizon with \emph{perfect} information and season-cyclic storage; hosting recourse
			\State $\mathrm{M2}$: joint community problem with pool \eqref{eq_pool}, perfect information, hosting by recourse (no price)
			\State $\mathrm{M3}$: joint community problem with \eqref{eq_pool} and the net-export cap internalized; set $(\pi_t,\tau_t)\gets(\hat\lambda^{\mathrm{pool}}_t,\hat\lambda^{\mathrm{host}}_t)$
			\For{$v\in\{\mathrm{DE0},\mathrm{M1},\mathrm{IND},\mathrm{M2},\mathrm{M3}\}$}
			
			\State $\Gamma_v\gets\mathrm{cost}(v)-\mathrm{cost}(\mathrm{SP})$;\quad $\varrho_v\gets(\Gamma_{\mathrm{DE0}}-\Gamma_v)/\Gamma_{\mathrm{DE0}}$
			\EndFor
			\For{$v\in\{\mathrm{M2},\mathrm{M3}\}$}
			\State $\varrho^{\mathrm{IND}}_v\gets(\Gamma_{\mathrm{IND}}-\Gamma_v)/\Gamma_{\mathrm{IND}}$
			\EndFor
			\State \Return KPIs, prices $(\pi_t,\tau_t)$, and $\{\Gamma_v,\varrho_v,\varrho^{\mathrm{IND}}_v\}$
		\end{algorithmic}
	\end{algorithm}
	By \Cref{prop_welfare}, the joint problems already yield the equilibrium allocation and the economic prices, which makes the master procedure exact and removes any need for price iteration. A distributed price-discovery implementation, in which agents best-respond to broadcast prices that a coordinator updates by a subgradient step on the raw coupling multipliers, is possible in principle for a platform that does not observe the agents' private constraints. It is not required for any presented result and is not developed in this paper.
	
	\section{Testbed, data, and experimental design}\label{sec_setup}
	
	\subsection{The district configurations}\label{subsec_testbed}
	
\begin{figure}
		\centering
		\includegraphics[width=0.7\textwidth]{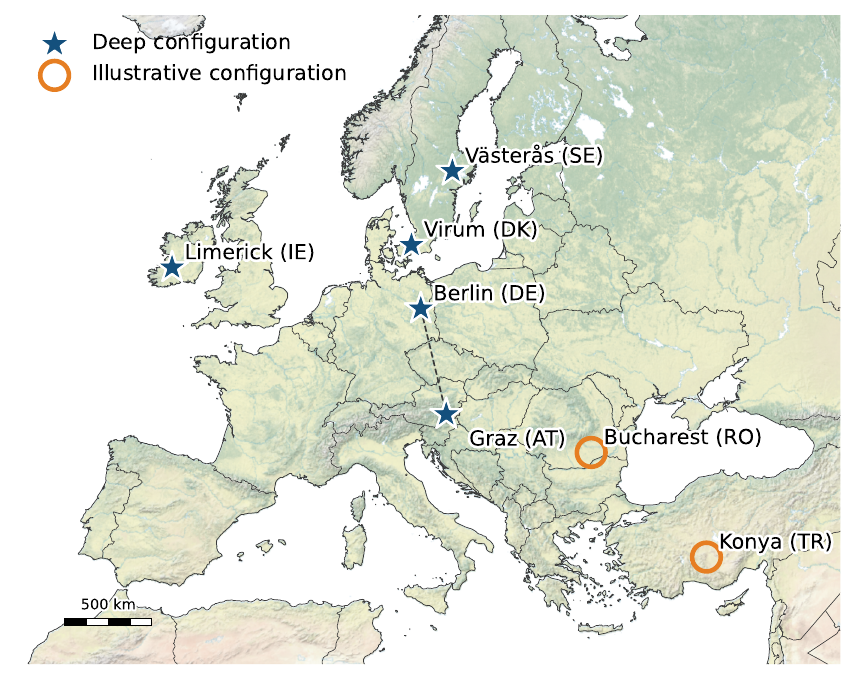}
		\caption{Six European district configurations, deep sites as stars and illustrative sites as circles, with the dashed link joining the Graz and Berlin demonstrators that merge into one power-to-heat archetype on the German zone.}
		\label{fig_geomap}
	\end{figure}

	The testbed comprises six European district configurations, four treated in depth and two as illustrative cases, spanning Nordic, central-European, and southern-European climate and market conditions. The power-to-heat configuration is a single composite archetype drawing on two demonstrator sites, Graz in Austria and Berlin in Germany, and is assigned the German bidding zone. It is not a physically aggregated cross-border district. The map shows seven site markers for the six configurations. Each configuration is a small community of heterogeneous prosumer agents whose renewable portfolio is sized to the positive energy threshold, making the district approximately annually balanced, the regime in which the hourly balancing question is meaningful. \Cref{fig_geomap} maps the sites, \Cref{tab_pilots} gives the aggregate scale, and \Cref{tab_agents} in \ref{app_data} the per-agent composition of the deep configurations. The annual electricity-equivalent demand $\mathcal{D}$ converts heat at the average heat-pump coefficient of performance, and the positivity ratio is renewable generation over that demand, RES$/\mathcal{D}$.
	\begin{table}[H]
		\centering
		\caption{The six district configurations and their annual scale.}
		\label{tab_pilots}
		\setlength{\tabcolsep}{4pt}
		\resizebox{\ifdim\width>\linewidth\linewidth\else\width\fi}{!}{%
			\begin{tabular}{@{}lcrrrrrrrr@{}}
				\toprule
				Configuration & Zone & Agents & PV (kWp) & RES (MWh) & Electricity (MWh) & Heat (MWh) & $\mathcal{D}$ (MWh) & RES$/\mathcal{D}$ & Host. (kW) \\
				\midrule
				Virum & DK2 & 4 & 1053 & 1032 & 743 & 813 & 983 & 1.05 & 180 \\
				Limerick & IE & 6 & 1454 & 1381 & 1079 & 618 & 1256 & 1.10 & 260 \\
				V\"aster\aa s & SE3 & 4 & 1506 & 1431 & 998 & 760 & 1363 & 1.05 & 160 \\
				Graz--Berlin & DE & 4 & 1045 & 1097 & 859 & 653 & 1045 & 1.05 & 200 \\
				Türkiye & TR & 3 & 313 & 485 & 406 & 98 & 422 & 1.15 & 160 \\
				Romania & RO & 3 & 681 & 886 & 715 & 308 & 886 & 1.00 & 130 \\
				\bottomrule
		\end{tabular}}
	\end{table}
	
	The \emph{Virum} configuration in Denmark is a sector-coupled social-housing district in the low-carbon, price-volatile Danish DK2 zone, and is the primary case for the sector-coupling and full-year analysis. The \emph{Limerick} configuration in Ireland is a mixed-use commercial-residential district and the primary P2P market case, its scale informed by a community-trading demonstrator in the Irish single-electricity market. The \emph{V\"aster\aa s} configuration in Sweden is an office district with photovoltaic-facade generation and strong heating, ventilation, and air-conditioning (HVAC) flexibility in the low-carbon Swedish SE3 zone. The \emph{Graz--Berlin} configuration is a composite power-to-heat archetype on a district-heating network with large thermal storage, in which electricity is converted to heat to absorb surplus. It draws on an Austrian demonstrator in Graz and a German demonstrator in Berlin, and takes the German bidding zone, whose power-to-heat calibration comes from the Berlin site. The two illustrative configurations broaden the climatic and technological range. The \emph{Türkiye} configuration is a high-irradiance, solar-assisted heat-pump district with photovoltaic-thermal collectors, where the mismatch between abundant summer generation and modest heat demand is largest. The \emph{Romania} configuration is a smart-HVAC office district on a district-heating network. They are presented in the aggregate results but not in the detailed decomposition and sensitivity analysis.
	
	\subsection{Data calibration and experimental design}\label{subsec_calibration}
	
	The hourly profiles are calibrated archetypes. Results computed from them are scenario-based projections. They are constructed by deterministic seasonal models whose annual totals match documented country values and the design parameters of each archetype, and every series is reproducible from a fixed seed. The horizon is four representative weeks, one per season, at hourly resolution, and each step $t$ carries an annualization weight $w_t=\tfrac{n_{\sigma(t)}}{H_\sigma/24}$, where $n_\sigma$ is the number of days the season represents, 90, 92, 92 and 91 for winter, spring, summer and autumn, and $H_\sigma$ the number of hours in the representative week, which scales representative-week sums to annual totals. A full-year, 8736-hour run (fifty-two weeks) is used on the primary configuration as a robustness check. Ambient temperature combines a seasonal mean, a diurnal cycle, and a stochastic disturbance,
	\begin{equation}\label{eq_tamb}
		T^{\mathrm{amb}}_t=\bar T_{\sigma(t)}+A_T\cos\!\Big(\tfrac{2\pi(h_t-15)}{24}\Big)+\varepsilon_t,
	\end{equation}
	where $\bar T_\sigma$ is the seasonal mean for the country, $h_t$ the hour of day, $A_T$ the diurnal amplitude with the daily minimum near dawn and the maximum in mid-afternoon, and $\varepsilon_t$ a small zero-mean disturbance. Solar availability follows a clear-sky daily bell over the seasonal daylight window, modulated by a clearness draw,
	\begin{equation}\label{eq_solar}
		g_t=\cos^2\!\Big(\tfrac{\pi}{2}x_t\Big)\,\mathbf{1}[|x_t|<1]\,S_{\sigma(t)}\,\zeta_t,\qquad x_t=\frac{h_t-12}{L_{\sigma(t)}/2},
	\end{equation}
	where $L_\sigma$ is the seasonal daylight length, $S_\sigma$ the seasonal solar strength, and $\zeta_t$ a bounded clearness factor. Photovoltaic availability of agent $a$ is the normalized solar profile scaled so that its annual energy matches the installed capacity times the country-specific yield,
	\begin{equation}\label{eq_pvscale}
		\mathrm{PV}^{\mathrm{av}}{at}=\phi_a,g_t,\qquad \phi_a\ \text{such that}\ \sum_t \mathrm{PV}^{\mathrm{av}}{at},w_t\Delta t = \mathrm{kWp}_a,Y_c,
	\end{equation}
	where $g_t$ is the dimensionless clear-sky solar profile of the site at step $t$, $\phi_a$ is the scaling factor of agent $a$ in kilowatts, $\mathrm{kWp}_a$ is the installed photovoltaic capacity of agent $a$ in kilowatt peak, $Y_c$ is the annual specific yield of country $c$ in kilowatt hours per kilowatt peak, and $w_t\Delta t$ is the annualization weight of step $t$ in hours. Photovoltaic-thermal heat is proportional to the same solar profile. Electrical load is built from archetype daily shapes for residential, commercial, office, and public buildings,
	\begin{equation}\label{eq_load}
		L_{at}=\bar P^{\mathrm{arch}}_a\,\psi_{a}(h_t)\,\chi_{\sigma(t)}\,(1+\nu_t),
	\end{equation}
	where $\bar P^{\mathrm{arch}}_a$ is the archetype mean power, $\psi_a$ a normalized daily shape combining a base level with morning, midday, and evening components weighted by the archetype occupancy pattern, $\chi_\sigma$ a seasonal scaling that raises consumption in the darker seasons, and $\nu_t$ a small disturbance. A fixed fraction of the load is treated as shiftable, $D^{\mathrm{sh}}_{at}=0.25\,L_{at}$, and the remainder as fixed, $D^{\mathrm{fix}}_{at}=0.75\,L_{at}$. Heat demand follows a degree-hour model sized by the design load density and modulated by occupancy,
	\begin{equation}\label{eq_heat}
		D^h_{at}=q^{\mathrm{des}}_a\,\Big[\tfrac{T^{\mathrm{base}}-T^{\mathrm{amb}}_t}{\Delta T^{\mathrm{des}}}\Big]^+_{\le 1.2}\,o_{at},\qquad q^{\mathrm{des}}_a=\iota_a F_a,
	\end{equation}
	where $q^{\mathrm{des}}_a$ is the design heat load, $\iota_a$ the heat load density, $F_a$ the floor area, $T^{\mathrm{base}}$ the heating base temperature, $\Delta T^{\mathrm{des}}$ the design temperature difference, the bracket a degree-hour ratio clipped to a maximum, and $o_{at}$ an occupancy modulation. Renewable capacity is not fixed a priori. It is sized such that each district reaches the positive energy threshold. The annual electricity-equivalent demand of the community is
	\begin{equation}\label{eq_demeq}
		\mathcal{D}=\sum_a\Big[\sum_t (D^{\mathrm{fix}}_{at}+D^{\mathrm{sh}}_{at})w_t\Delta t + \frac{\sum_t (D^h_{at}-Q^{\mathrm{pvt}}_{at})^+ w_t\Delta t}{\overline{\mathrm{COP}}_a}\Big],
	\end{equation}
	where $\overline{\mathrm{COP}}_a$ is the annual average heat-pump coefficient of performance of agent $a$, at which the heat demand is converted to electricity for heat-pump agents and at unit efficiency otherwise. The target annual renewable generation is $r^{\mathrm{ped}}\mathcal{D}$, where $r^{\mathrm{ped}}$ is the positive energy ratio, and it is allocated across agents in proportion to roof area $F_a$, which sets each agent's installed capacity $\mathrm{kWp}_a=r^{\mathrm{ped}}\mathcal{D}\,(F_a/\sum_b F_b)/Y_c$ used in \eqref{eq_pvscale}. This procedure places every district at, or a chosen fraction above, annual balance, which makes the hourly balancing question well posed, and it is the sizing that the renewable-penetration sensitivity varies.
	
	Import prices combine a zone mean with a diurnal shape, calibrated to zonal day-ahead ranges on the ENTSO-E Transparency Platform \citep{entsoe}. Export prices are a zone-specific fraction of the import price, a volumetric use-of-system tariff applies in both directions as a stylized symmetric network charge, with the export-side charge a modeling assumption that \Cref{subsec_res_sens} sweeps to zero, and the emission factors follow the same zone-mean-and-diurnal-shape form. These emission factors are average, location-based zonal intensities, not short-run marginal or consequential factors, and the presented emissions are attributional. The remaining zone, technology, and flexibility parameters are given in \ref{app_data}, including the country-specific photovoltaic yields calibrated to PVGIS climate ranges \citep{pvgis}. Each configuration's technology set, scale, and design anchors (installed capacities, self-consumption improvement, P2P price spreads, and power-to-heat peak reduction) are drawn from the corresponding demonstrator's public documentation, whereas every hourly series is generated deterministically and none is metered. Each parameter is a documented archetype value whose status and source type appear in \Cref{tab_tech}. Validation against measured pilot data or open datasets is left to future work, and the results should be read accordingly.
	
	The fabric sensitivity needs two parameters and one derived rate. The effective heat capacity is $c^{\mathrm{f}}=0.08$~kWh per square meter per kelvin, between the heavy and very heavy internal heat capacity classes, which suits the masonry blocks and the public building of the primary configuration. The comfort band half-width $\Delta\theta$ is the sweep variable. The standing-loss rate needs no new assumption, as the envelope conductance follows from parameters already in the model as $UA_a=q^{\mathrm{des}}_a/\Delta T^{\mathrm{des}}$ and the fabric capacity is $c^{\mathrm{f}}F_a$, giving $\kappa^{\mathrm{f}}_a=UA_a/(c^{\mathrm{f}}F_a)$, the inverse building time constant. That is 1.88\% per hour for the residential blocks and 1.67\% for the public building, time constants of 53 and 60 hours. The characterization of fabric flexibility follows \citet{reynders2017generic}. The net-export limit is physically enforced by a feasible recourse even in the variants where it is not priced, the independent variants and the pooled variant without grid-aware pricing. After each agent's schedule is realized, the DSO throttles the community net export at the substation to $\bar E_t$ in any hour where it would otherwise exceed the limit. The throttling curtails dispatched photovoltaic power and reduces the matching export equally, preserving every agent's electricity balance \eqref{eq_ebal}. Since the curtailment cannot exceed the dispatched power, the realized schedule stays feasible. The curtailed energy is recorded as renewable spill and attributed to exporting agents in proportion to their export, allocating both the lost generation and the lost revenue. This pro rata rule is one of several admissible conventions, alongside allocation by installed capacity or a uniform share, and the agent-level outcomes depend on the choice even though the community totals do not. It is a distributional convention for a settlement design to fix, separate from the aggregate results we present. The residual overflow beyond the curtailable photovoltaic power is zero to four decimals in every configuration. The recourse is a genuine feasibility restoration, not an accounting truncation.
	
	The experiments combine a core run matrix with three complementary families of tests. For each of the six configurations, the core matrix evaluates the mixed-integer community operator $\mathrm{SP}$, its convex relaxation $\mathrm{SP}_{\mathrm{rel}}$, and the five decentralized stages $\mathrm{DE0}$, $\mathrm{M1}$, $\mathrm{IND}$, $\mathrm{M2}$, and $\mathrm{M3}$. It identifies the residual bidirectional hourly gap, the no-platform cost gap, the ordered marginal contributions along the chosen waterfall, and the P2P and hosting prices. $\mathrm{M1}$ is the shared-forecast step. $\mathrm{IND}$, $\mathrm{M2}$, and $\mathrm{M3}$ are the perfect-information stages that bound what organization and grid pricing can deliver within the convex model.
	The second family constructs a self-sufficiency cost curve by tightening an annual-import bound in the community problem. The third varies eight physical and economic drivers one at a time, namely the hosting limit, carbon price, storage capacity, renewable penetration, demand-response depth, and feed-in price, together with a bounded within-day fabric store and an export-side use-of-system tariff. We also perform two independent convex-solver checks on the primary configuration. A regularization sensitivity over $\varepsilon^{\mathrm r}\in\{10^{-7},10^{-6},10^{-5}\}$ confirms that the presented values are not artifacts of the coupling-flow regularizer, and an alternate-order attribution, introducing P2P trading and grid-aware pricing in both possible orders, confirms that they do not depend on the chosen sequential path.
	
	\section{Results and analysis}\label{sec_results}
	
	The complete per-configuration and per-variant results are provided in the reproduction package, and \ref{app_extra} presents the supporting checks and decompositions. All presented values are scenario-based projections computed from the calibrated archetypes.
	
	\subsection{The residual hourly imbalance under central coordination}\label{subsec_res_size}
	
\begin{figure}[H]
		\centering
		\includegraphics[width=1\textwidth]{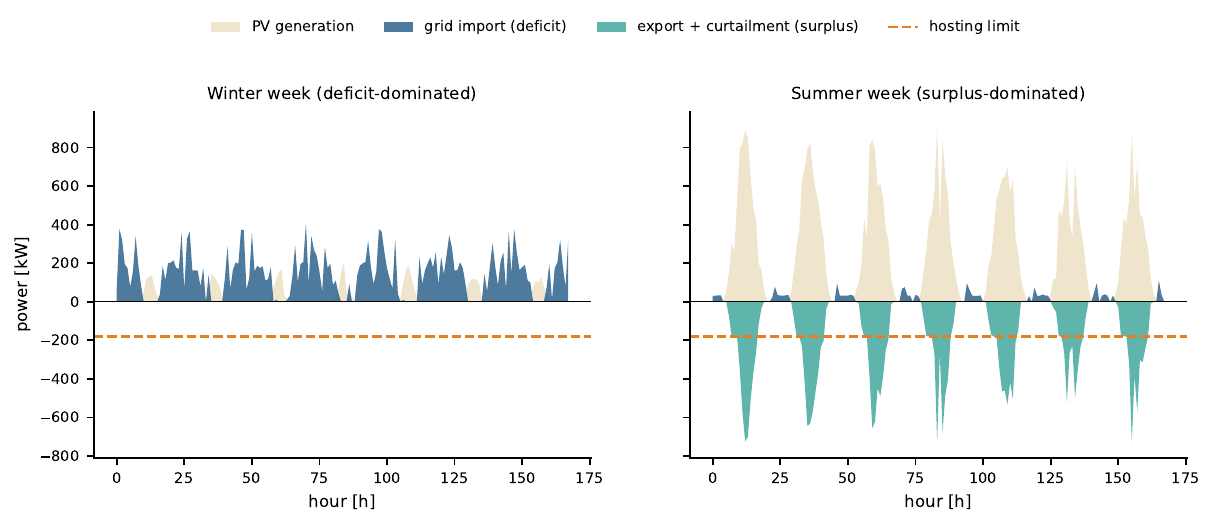}
		\caption{Hourly grid import, export plus curtailment, and PV generation over a winter (left) and a summer (right) representative week for an annually positive district (Virum, planner).}
		\label{fig_concept}
	\end{figure}

	\Cref{fig_concept} illustrates the problem for the Virum configuration. The district is annually positive, though a winter representative week is deficit-dominated, with heavy import and little generation, while a summer representative week is surplus-dominated, with generation exceeding demand and export clipped at the hosting limit, which appears as curtailment. The same district imports through winter nights and spills renewable output through summer middays, two burdens in different seasons and directions.
	
	\Cref{tab_main} gives the planner benchmark for all six configurations. Even under perfect central coordination the self-sufficiency stays near half, 48 to 50\% across the four deep configurations and up to 55\% for the illustrative smart-HVAC configuration. The hourly matching rate is 36 to 47\%, and 46 to 53\% of renewable output is exported or curtailed. Therefore, annual positivity leaves a large residual hourly imbalance in both directions. The oversupply component depends strongly on the hosting limit and the local demand structure. The Nordic and central-European districts, which have district heating and stronger sector coupling, reach comparable self-sufficiency at lower oversupply than the high-irradiance illustrative case, where the summer surplus is the largest share of a modest generation base. That case nonetheless curtails little and binds in the fewest hours, as \Cref{tab_prices} shows. Its hosting headroom is generous relative to its smaller installed capacity, and a high oversupply fraction need not imply frequent binding or heavy curtailment. In every studied run the community net export does not exceed the contemporaneous renewable surplus, with the coverage share equal to one to four decimal places, which supports attributing net export to renewable generation in the oversupply accounting. The last column of \Cref{tab_main} gives the coordination gap, the cost distance from the no-platform decentralized outcome to the planner, which is the subject of the next subsection.
	
	\begin{table}[H]
		\centering
		\caption{Planner benchmark per configuration, presenting cost, emissions, hourly self-sufficiency, oversupply, import, curtailment, and the coordination gap.}
		\label{tab_main}
		\setlength{\tabcolsep}{4pt}
		\resizebox{\ifdim\width>\linewidth\linewidth\else\width\fi}{!}{%
			\begin{tabular}{@{}lrrcccrrrr@{}}
				\toprule
				District & Cost (k\euro) & CO$_2$ (t) & SSR & HMR & LPI$_{os}$ & Import (MWh) & Curtail (MWh) & Gap (k\euro) & Gap (\% of DE0) \\
				\midrule
				Virum & 50.7 & 43.3 & 0.50 & 0.47 & 0.46 & 565 & 165 & 3.17 & 5.9\% \\
				Limerick & 91.3 & 177.9 & 0.49 & 0.38 & 0.53 & 676 & 255 & 4.87 & 5.1\% \\
				V\"aster\aa s & 60.4 & 19.0 & 0.48 & 0.37 & 0.53 & 728 & 433 & 1.75 & 2.8\% \\
				Graz--Berlin & 84.0 & 181.4 & 0.48 & 0.36 & 0.47 & 612 & 163 & 3.83 & 4.4\% \\
				Türkiye & 23.5 & 79.7 & 0.53 & 0.39 & 0.50 & 212 & 7 & 1.24 & 5.0\% \\
				Romania & 51.8 & 110.1 & 0.55 & 0.37 & 0.49 & 375 & 185 & 1.23 & 2.3\% \\
				\bottomrule
		\end{tabular}}
	\end{table}
	
\begin{figure}[H]
		\centering
		\includegraphics[width=\textwidth]{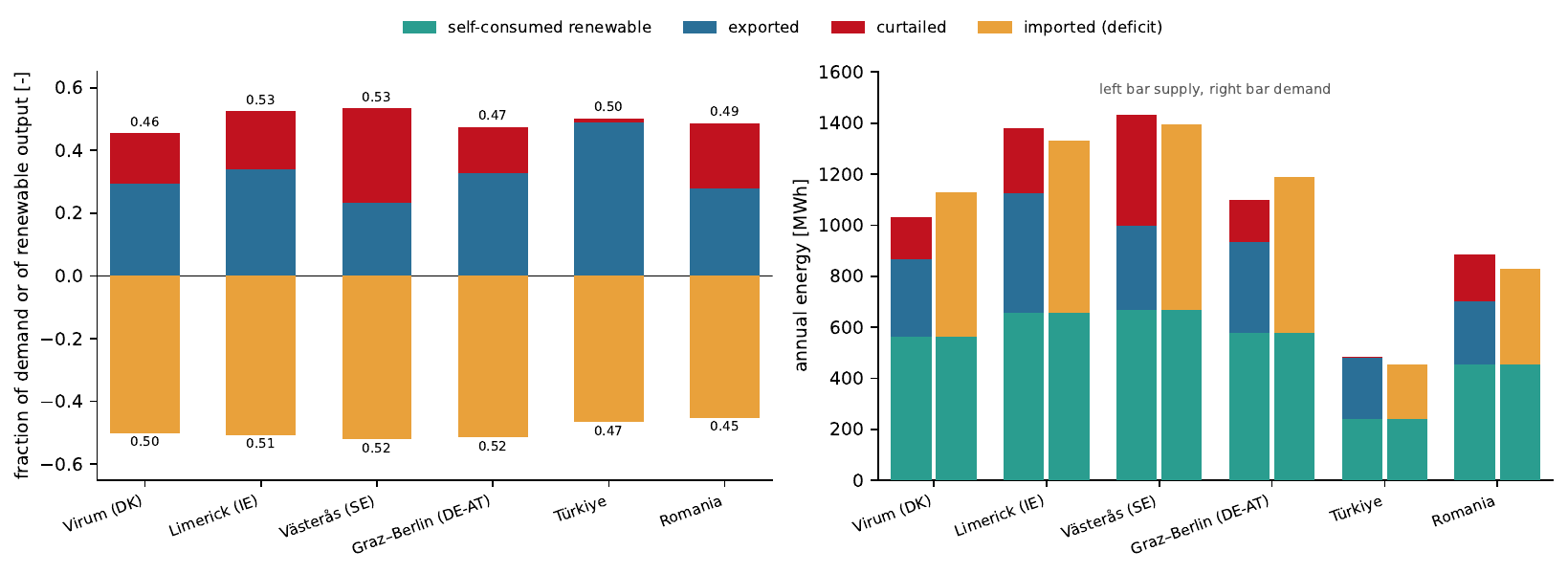}
		\caption{Residual hourly imbalance per configuration under the planner, as shares of demand and of renewable output (left) and on an absolute annual energy scale (right).}
		\label{fig_last20}
	\end{figure}

\begin{figure}[H]
		\centering
		\includegraphics[width=0.8\textwidth]{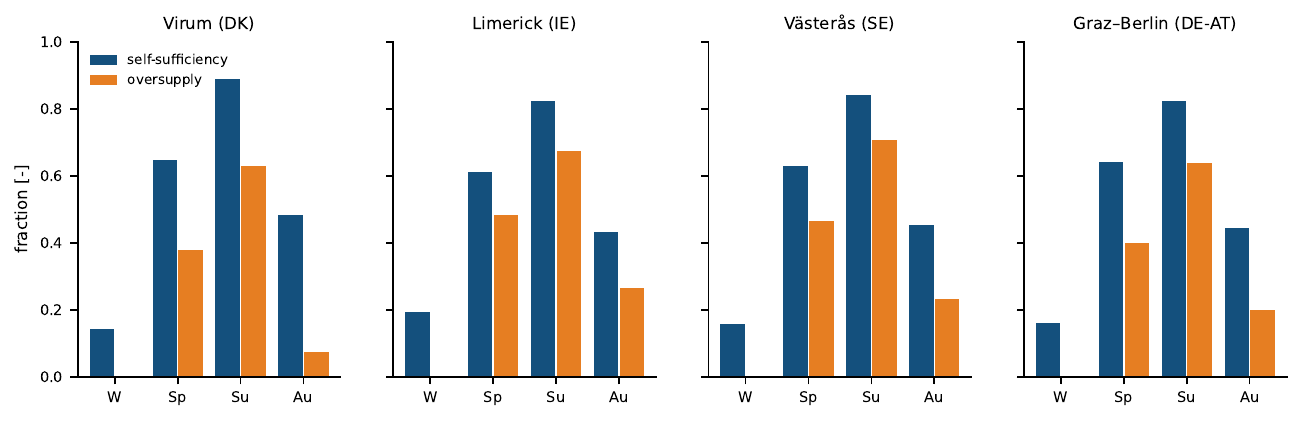}
		\caption{Hourly self-sufficiency and oversupply by season (winter, spring, summer, autumn) for the deep configurations under the planner.}
		\label{fig_seasonal}
	\end{figure}

	\Cref{fig_last20} shows the residual hourly imbalance for all six configurations under the planner. The deficit reaches 45 to 52\% of demand and the oversupply 46 to 53\% of renewable output. Its right panel places these fractions on an absolute energy scale, decomposing each district's renewable output into self-consumed, exported, and curtailed energy, and its demand into the self-consumed renewable share and imported energy.
	\Cref{fig_seasonal} locates the gap in time. Winter is deficit-dominated, with self-sufficiency near 0.2 and almost no oversupply, whereas summer reaches self-sufficiency near 0.9 with substantial oversupply. The annual balance conceals a strong seasonal alternation between the two burdens, which is the operational core of the residual hourly imbalance.
	
	\subsection{The coordination gap and the ordered mechanism values}\label{subsec_res_gap}
	
\begin{figure}[htbp]
		\centering
		\includegraphics[width=0.9\textwidth]{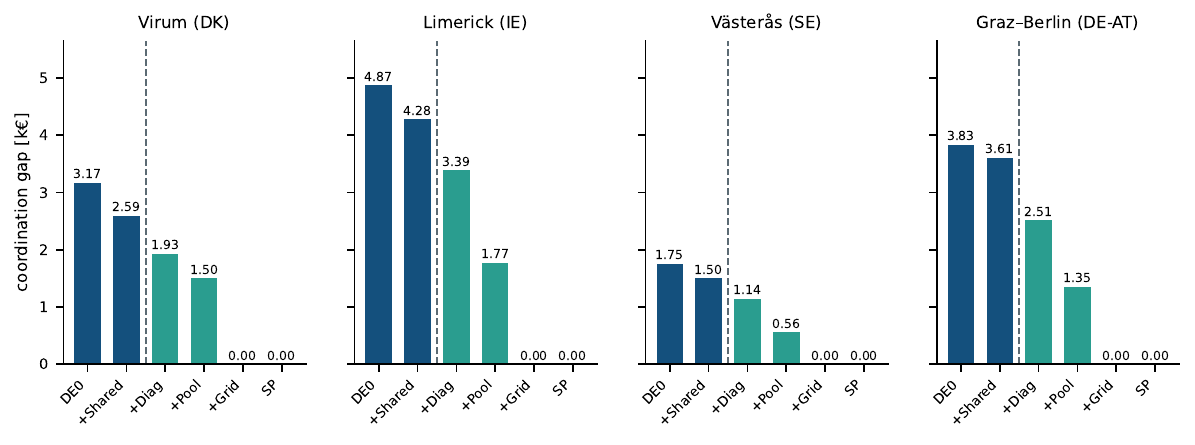}
		\caption{
			Coordination gap by stage for the deep configurations, with the dashed line separating the forecast-informed experiment from the perfect-information one and the gap across it unattributed.}
		\label{fig_waterfall}
	\end{figure}

	\Cref{fig_waterfall} and \Cref{tab_waterfall} show the no-platform gap and its ordered attribution. The difference between $\mathrm{DE0}$ and the community operator is 2--6\% of annual economic cost across the six configurations. Replacing private forecasts with the shared forecast in $\mathrm{M1}$, the only forecast-informed operational stage, recovers 6 to 18\% of that gap under the same daily rolling implementation with perfect intraday recourse. That is the whole of the first experiment. The transition $\mathrm{M1}\!\to\!\mathrm{IND}$ is the unattributed residual between the two experiments. The second experiment starts from $\mathrm{IND}$ and measures its two mechanisms against the $\mathrm{IND}$-to-$\mathrm{SP}$ distance. The step $\mathrm{IND}\!\to\!\mathrm{M2}$ isolates the organizational value of P2P trading, which closes 23\% of that distance at Virum and 33 to 64\% elsewhere. The final step $\mathrm{M2}\!\to\!\mathrm{M3}$ internalizes the hosting constraint and takes the remainder, which is the larger share in four of the six configurations and the smaller share in the two more complementary ones.
	
	The full recovery of the gap by $\mathrm{M3}$ combines a duality result with a numerical finding. It is not an experimental outcome. By \Cref{prop_welfare}, conditional on the convex price-taking model and common perfect information, P2P clearing plus the hosting price implements the regularized planner relaxation $\mathrm{SP}_{\mathrm{rel}}$, which makes $\mathrm{M3}=\mathrm{SP}_{\mathrm{rel}}$ by construction. It additionally equals the mixed-integer planner $\mathrm{SP}$ and closes the $\mathrm{DE0}$-to-$\mathrm{SP}$ gap in full in the studied instances, since the observed integrality gap $\mathrm{SP}-\mathrm{SP}_{\mathrm{rel}}$ is 0.0000 thousand euro in every configuration. The directly observed, forecast-informed contribution is the $\mathrm{DE0}\!\to\!\mathrm{M1}$ reduction of 6 to 18\%. The remaining perfect-information stages identify an upper bound on the organizational and tariff value that better forecasting and implementation could approach.
	
	The mechanisms interact, which makes the waterfall path-dependent. An independent cross-check introduces P2P trading and grid pricing in both orders for Virum, against an $\mathrm{IND}$-to-$\mathrm{SP}$ distance of 1.934 k\euro. Taking the pool first attributes 0.435 k\euro{} (22.5\%) to the pool and 1.499 k\euro{} (77.5\%) to grid-aware pricing, and taking the grid price first attributes 0.491 k\euro{} (25.4\%) and 1.444 k\euro{} (74.6\%). The two-order average is 0.463 k\euro{} (23.9\%) for the pool and 1.471 k\euro{} (76.1\%) for grid-aware pricing, with an interaction of 0.055 k\euro{} (2.9\%), which makes the two mechanisms mildly complementary. The qualitative conclusion for Virum, that grid-aware pricing is the larger contributor while P2P trading has a positive though smaller role, does not depend on the chosen order. In V\"aster\aa s and Türkiye the pool is the larger, taking 51 and 64\% of the perfect-information distance, their hosting limits binding in fewer hours. Measuring both mechanisms against the $\mathrm{IND}$-to-$\mathrm{SP}$ distance, which is the baseline of the experiment that contains them, leaves that ordering unchanged for Virum. The regularization check likewise leaves the economic cost and qualitative ranking stable over two orders of magnitude in $\varepsilon^{\mathrm r}$.
	
	\begin{table}[H]
		\centering
		\caption{Annual economic cost by stage and configuration, the coordination gap $\Gamma_{\mathrm{DE0}}$, the share of that gap closed by the shared-forecast step, $\varrho_{\mathrm{M1}}$, and the share of the $\mathrm{IND}$-to-$\mathrm{SP}$ distance closed by the pool, $\varrho^{\mathrm{IND}}_{\mathrm{M2}}$, the grid price taking the remainder.}
		\label{tab_waterfall}
		\setlength{\tabcolsep}{4pt}
		\resizebox{\ifdim\width>\linewidth\linewidth\else\width\fi}{!}{%
			\begin{tabular}{@{}lrrrrrrrrr@{}}
				\toprule
				District & DE0 (k\euro) & M1 (k\euro) & IND (k\euro) & M2 (k\euro) & M3 (k\euro) & SP (k\euro) & $\Gamma_{\mathrm{DE0}}$ (k\euro) & $\varrho_{\mathrm{M1}}$ (\%) & $\varrho^{\mathrm{IND}}_{\mathrm{M2}}$ (\%) \\
				\midrule
				Virum & 53.8 & 53.2 & 52.6 & 52.2 & 50.7 & 50.7 & 3.17 & 18 & 23 \\
				Limerick & 96.2 & 95.6 & 94.7 & 93.1 & 91.3 & 91.3 & 4.87 & 12 & 48 \\
				V\"aster\aa s & 62.1 & 61.9 & 61.5 & 60.9 & 60.4 & 60.4 & 1.75 & 14 & 51 \\
				Graz--Berlin & 87.9 & 87.6 & 86.5 & 85.4 & 84.0 & 84.0 & 3.83 & 6 & 46 \\
				T\"urkiye & 24.7 & 24.6 & 24.1 & 23.7 & 23.5 & 23.5 & 1.24 & 14 & 64 \\
				Romania & 53.1 & 52.9 & 52.7 & 52.4 & 51.8 & 51.8 & 1.23 & 14 & 33 \\
				\bottomrule
		\end{tabular}}
	\end{table}
	
	The physical indicators show a complementary pattern. Shared information and internal trading improve the use of existing flexibility, while the hosting price primarily reduces surplus-side curtailment. The deficit-side self-sufficiency remains bounded by installed storage and demand flexibility. Therefore, coordination closes an avoidable operating-cost gap and improves export management, though it does not substitute for the physical capacity needed to shift energy across deficit hours.
	
	\subsection{Equilibrium prices}\label{subsec_res_prices}
	
\begin{figure}[H]
		\centering
		\includegraphics[width=0.9\textwidth]{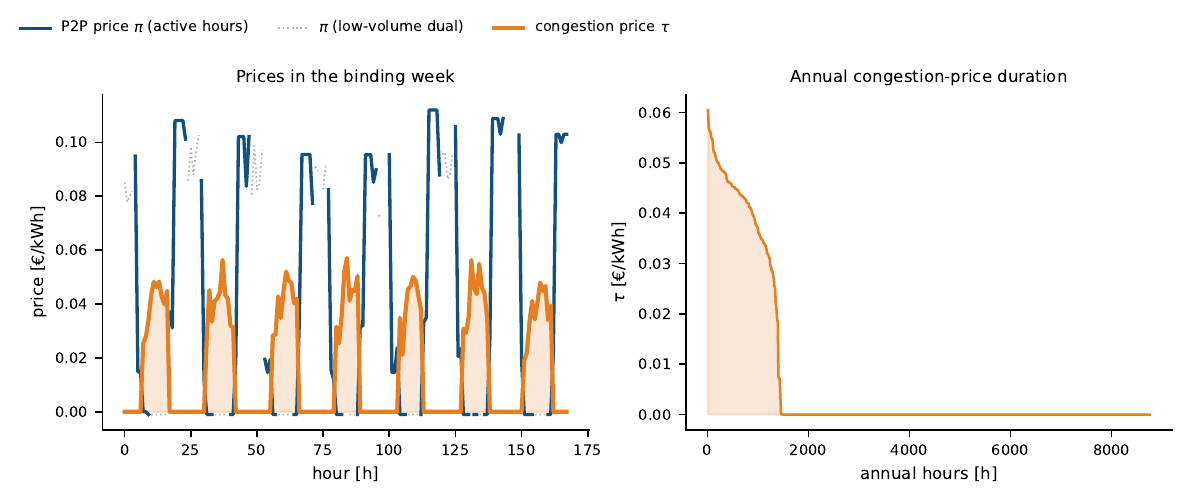}
		\caption{P2P and congestion prices over the binding representative week on the left, solid in active pool hours ($V_t>1$~kW) and greyed dotted in the low-volume hours where the dual is solver-selected, and the weight-expanded annual congestion-price duration curve for Virum on the right.}
		\label{fig_prices}
	\end{figure}

	\Cref{tab_prices} and \Cref{fig_prices} present the equilibrium prices, which are the dual multipliers of the pool balance and the net-export hosting limit by \Cref{prop_welfare}. The pool-balance multiplier is not unique in hours with negligible pool volume, which in turn help summarizing the P2P clearing price by its trade-volume-weighted mean over active hours. With aggregate pool volume $V_t=\sum_a s_{at}=\sum_a b_{at}$ and active set $\{t:V_t>1~\mathrm{kW}\}$ before annualization, this weighted price is $\bar\pi_{\mathrm{tw}}=\sum_t w_t V_t \pi_t/\sum_t w_t V_t$. The seller premium and buyer saving are the analogous $V_t$-weighted means of $\pi_t-r^{\mathrm{exp}}_t$ and $c^{\mathrm{imp}}_t-\pi_t$ against the economic-cost $\mathrm{M3}$ alternatives $r^{\mathrm{exp}}_t=\lambda^{\mathrm{exp}}_t-\tau^{\mathrm{use}}-\tau_t$ and $c^{\mathrm{imp}}_t=\lambda^{\mathrm{imp}}_t+\tau^{\mathrm{use}}+\mu\,\mathrm{EF}^{\mathrm{g}}_t-\tau_t$, which net the use-of-system tariff, the congestion price, and, on import, the monetized carbon charge and exclude the negligible regularizer derivative. The pool is active in about 7600 to 8400 hours per year across the deep configurations. The weighted price lies within the effective band, at about 0.07 to 0.14~\euro/kWh. Since a single clearing price implies different values on the two sides, the seller premium is 0.06 to 0.11~\euro/kWh and the buyer saving 0.03 to 0.10~\euro/kWh, both broadly consistent with the qualitative spreads discussed in the peer-to-peer market literature \citep{sousa2019peer,tushar2020peer}. The all-hour arithmetic mean of the multiplier, 0.07 to 0.13~\euro/kWh, is one particular solver dual. The congestion price is zero for most of the year and positive only when the net-export limit binds, on the order of 700 to 1800 hours per year, a tenth to a fifth of all hours, concentrated in the high-generation middays. Since the annual mean dilutes this signal, \Cref{tab_prices} also gives the mean congestion price conditional on binding, 0.02 to 0.05~\euro/kWh, and its annual maximum, 0.04 to 0.07~\euro/kWh. Its concentration in the high-generation hours and its low level when active make grid-aware pricing a targeted intervention. The congestion price is the shadow value of the one modeled cap. A network-constrained formulation would produce location-dependent prices and would redistribute curtailment across agents. The aggregate value of grid-aware pricing is a projection for the modeled interface, and the per-agent incidence is the part most sensitive to the simplification. In the modeled convex, price-taking setting, the duality lets the operator set a congestion charge that induces the coordinated net-export outcome without observing the community's internal decisions. The resulting congestion revenue, its incidence across prosumers, the ex ante rule, and two ex-post illustrations are presented in \Cref{tab_recycling}. A complete implementable settlement design, specifying revenue recycling, the charging of local trades for network use, and participant protection, remains future work.
	
	\begin{table}[H]
		\centering
		\caption{Equilibrium prices and market-power indicators per configuration, with the trade-weighting, the active pool set, and the economic-cost $\mathrm{M3}$ grid alternatives as defined in the text, and the Herfindahl index, largest-exporter share, and congestion exposure taken over the binding hours.}
		\label{tab_prices}
		\setlength{\tabcolsep}{4pt}
		\resizebox{\ifdim\width>\linewidth\linewidth\else\width\fi}{!}{%
			\begin{tabular}{@{}lrrrrrrrr@{}}
				\toprule
				District & $\bar\pi_{\mathrm{tw}}$ (\euro/kWh) & Pool active (h/yr) & $\bar\tau\,|\,\tau>0$ (\euro/kWh) & $\tau^{\max}$ (\euro/kWh) & Binding (h/yr) & HHI & Largest share & Exposure (\% of bill) \\
				\midrule
				Virum & 0.093 & 7579 & 0.040 & 0.061 & 1446 & 0.25 & 0.26 & 16.1 \\
				Limerick & 0.136 & 8366 & 0.046 & 0.068 & 1366 & 0.17 & 0.19 & 15.8 \\
				V\"aster\aa s & 0.067 & 7551 & 0.024 & 0.044 & 1759 & 0.25 & 0.27 & 12.5 \\
				Graz--Berlin & 0.117 & 8353 & 0.039 & 0.058 & 1419 & 0.25 & 0.28 & 13.4 \\
				Türkiye & 0.070 & 8642 & 0.023 & 0.036 & 683 & 0.35 & 0.44 & 8.8 \\
				Romania & 0.113 & 7319 & 0.035 & 0.050 & 1496 & 0.33 & 0.34 & 7.9 \\
				\bottomrule
		\end{tabular}}
	\end{table}
	
\begin{figure}[H]
		\centering
		\includegraphics[width=0.52\textwidth]{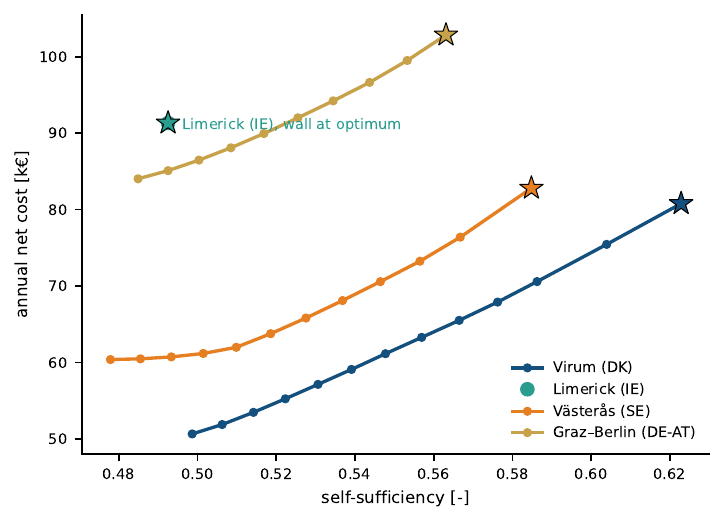}
		\caption{Annual net cost versus hourly self-sufficiency under a tightening planner import bound, one curve per configuration, with stars marking the maximum feasible self-sufficiency that Limerick reaches at its unconstrained optimum.}
		\label{fig_selfcost}
	\end{figure}

	\Cref{fig_selfcost} traces the cost of pushing the district toward higher hourly self-sufficiency, obtained by tightening a bound on annual import in the planner. The cost rises steeply and convexly as self-sufficiency increases, and each configuration reaches a feasibility wall beyond which the residual hourly imbalance cannot be closed with the installed storage and flexibility. The wall is closest for the mixed-use configuration with the highest and most commercial demand, whose wall lies just above its unconstrained optimum and leaves the smallest feasible headroom of the deep configurations, and furthest for the sector-coupled social-housing configuration. Coordination and the platform close the cost gap at low incremental resource cost and reduce oversupply. However, the deficit side is ultimately bounded by physical storage, and pushing self-sufficiency past the point that the installed storage supports requires investment, not better information or markets, at a cost that rises sharply and eventually becomes infeasible.
	
	\subsection{Sensitivity analysis}\label{subsec_res_sens}
	
\begin{figure}[H]
		\centering
		\includegraphics[width=0.8\textwidth]{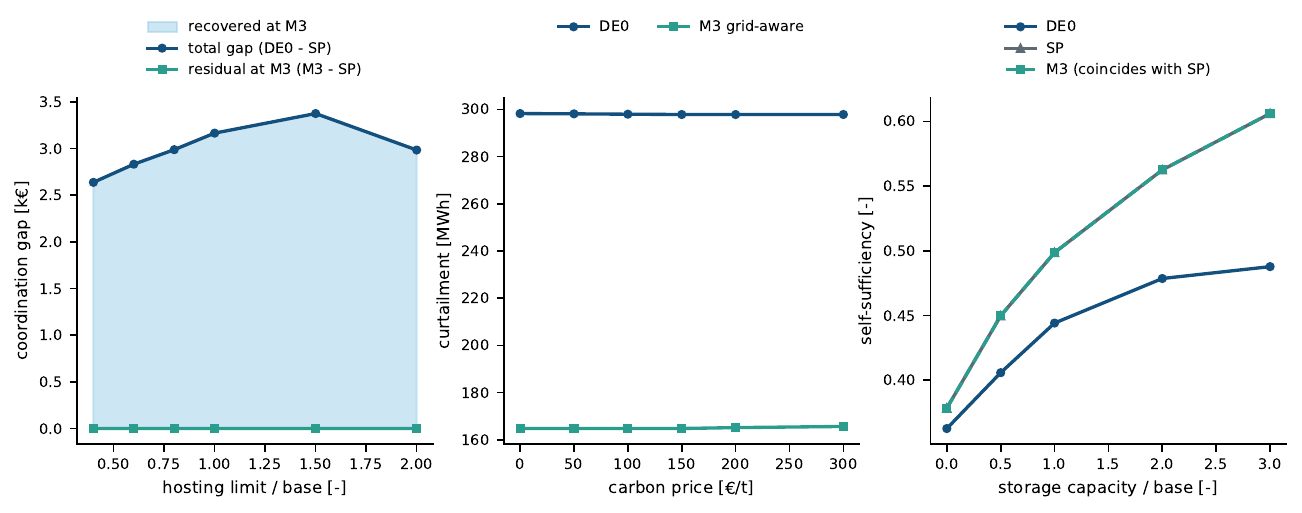}
		\caption{Sensitivity of the coordination gap (left), curtailment (center), and self-sufficiency (right) to the hosting limit, carbon price, and storage capacity for Virum, with $\mathrm{M3}$ drawn dashed in the right panel where it coincides with $\mathrm{SP}$.}
		\label{fig_sens1}
	\end{figure}

\begin{figure}[H]
		\centering
		\includegraphics[width=0.92\textwidth]{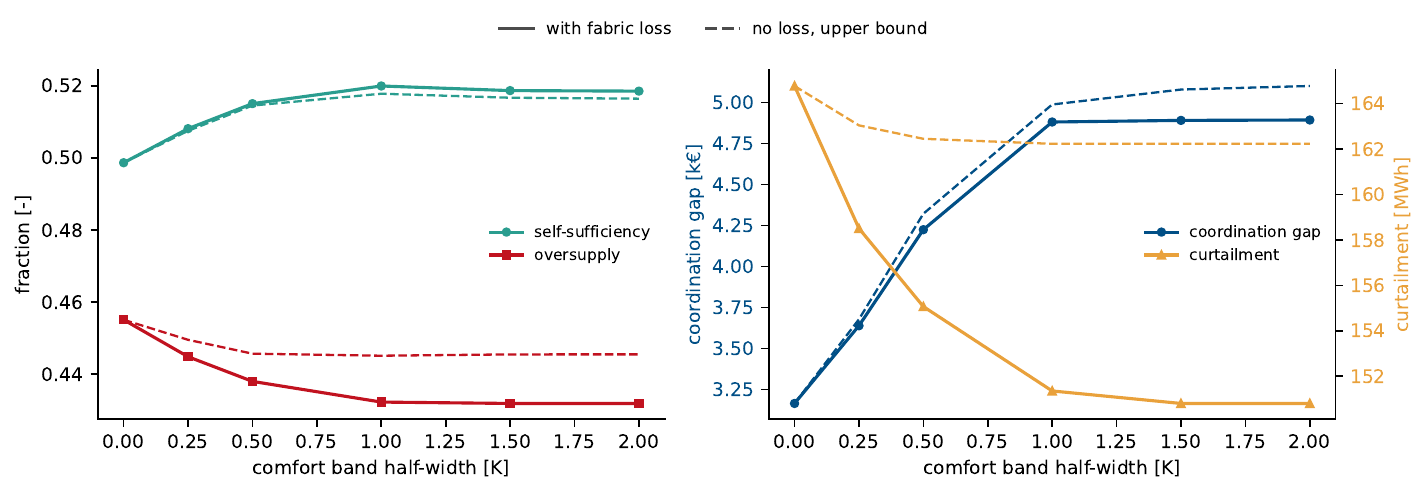}
		\caption{Thermal-mass sweep on Virum against the comfort band half-width, with the lossy fabric store solid and the lossless upper bound dashed.}
		\label{fig_thermalmass}
	\end{figure}

\begin{figure}[H]
		\centering
		\includegraphics[width=0.8\textwidth]{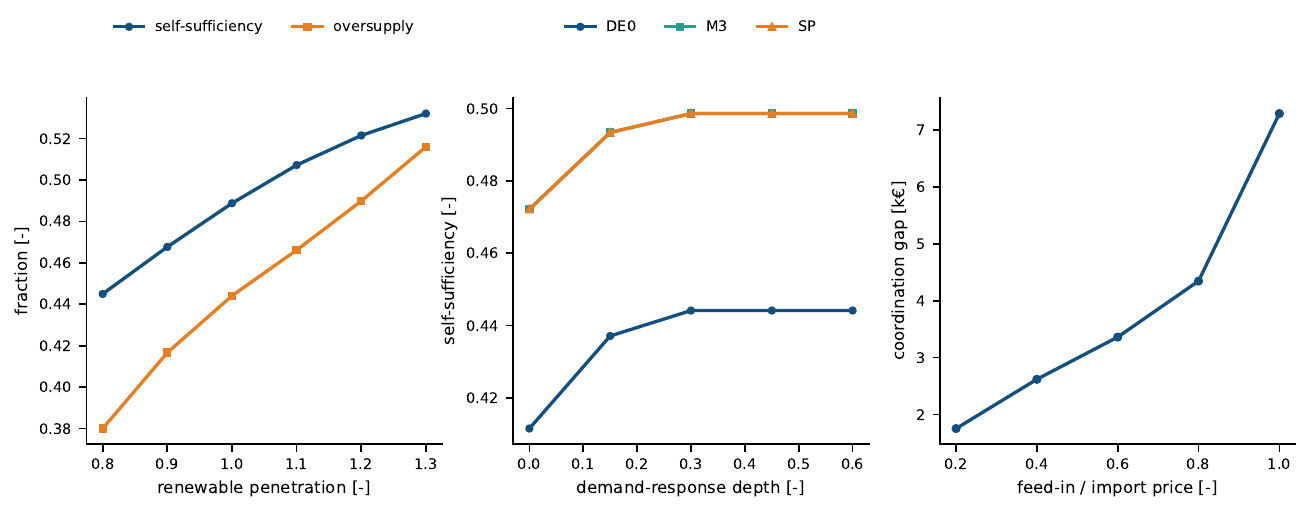}
		\caption{Sensitivity of self-sufficiency and oversupply (left), self-sufficiency (center), and the coordination gap (right) to renewable penetration, demand-response depth, and feed-in price (Virum).}
		\label{fig_sens2}
	\end{figure}

	Figures \ref{fig_sens1}-\ref{fig_sens2} present the sensitivities, using the Virum configuration as the reference.
	
	\emph{Grid scarcity:} The left panel of \Cref{fig_sens1} shows that the coordination gap is largest at intermediate grid scarcity. When the hosting limit is very loose, there is no congestion to mismanage, and the no-platform outcome is already near the planner. When the limit is very tight all organizations are equally constrained, and there is little room for coordination to help. In between, the gap peaks, and the perfect-information stages recover essentially the whole gap across the full range, with the residual after $\mathrm{M3}$ remaining near zero. Therefore, the value of coordination is greatest for the moderately constrained networks that are typical of urban distribution grids.
	
	\emph{Carbon price:} The center panel of \Cref{fig_sens1} shows that curtailment falls with grid-aware pricing at every carbon price, and that the reduction is roughly constant across the carbon range. This shows that the value of grid-aware pricing comes from the physical hosting limit, not from the carbon signal. The congestion price responds to the export cap. The carbon price shifts the level of emissions and the merit order between the heat pump and district heating, though it does not change the curtailment that the hosting limit drives.
	
	\emph{Storage:} The right panel of \Cref{fig_sens1} shows that self-sufficiency rises with storage under every organization, and that the ordering of planner, platform, and no-platform is preserved. Storage moves the deficit side of the residual hourly imbalance that coordination cannot, and it does this under any organization, which separates the role of storage from the role of coordination. The gap between the planner and the no-platform outcome widens with storage, from \euro{}670 to \euro7,480  across storage multiples zero to three, which confirms that storage and coordination are complementary.
	
	\emph{Renewable penetration:} The left panel of \Cref{fig_sens2} shows that higher renewable penetration raises oversupply faster than it raises self-sufficiency. Beyond the positive energy threshold the district exports and curtails more without becoming much more self-sufficient. The additional generation arrives in the summer middays that are already in surplus. This strengthens the case for storage and grid-aware pricing over simply adding generation, and it is a caution against reading a higher annual surplus as a better district.
	
	\emph{Demand-response depth:} The center panel of \Cref{fig_sens2} shows that deeper demand response raises self-sufficiency and reduces the gap, with diminishing returns. The saturation is set by the three-hour store capacity and the daily reset, not by the flexibility fraction $\alpha_a$ itself. Once $\alpha_a$ exceeds the $0.25$ shiftable share of load, an hour's entire shiftable load can already be moved, so the sweep beyond that point mainly varies the rebound rate and carries little additional information. Demand response complements storage and does not replace it.
	
	\emph{Feed-in price:} The right panel of \Cref{fig_sens2} shows that the coordination gap grows as the feed-in price rises toward the import price, from about two to seven thousand euro as the ratio moves from $0.2$ to $1.0$. When export is well remunerated, uncoordinated prosumers compete for the scarce export capacity capped by the shared hosting limit and lose revenue to curtailment. The coordinated community instead allocates that capacity efficiently and clears surplus internally through the pool, and its value rises with the price of the exports at stake. The platform is most valuable where export is well remunerated but grid capacity is scarce.
	
	\emph{Building thermal mass:} Figure \ref{fig_thermalmass} sweeps the comfort band half-width on the primary configuration, with the fabric store treated as a recourse decision in the no-platform variant exactly as the thermal tank is, with only the battery schedule committed. Hourly self-sufficiency rises from 0.499 to 0.520 and oversupply falls from 0.455 to 0.432, both saturating near a one-kelvin half-width. Curtailment falls from 165 to 151~MWh. The coordination gap does not shrink when thermal mass is added. It rises from \euro{}3,170 to \euro{}4,890 and from 5.9 to 9.4 \% of the no-platform cost. The no-platform cost itself falls only slightly, from \euro{}53,800 to \euro{}52k, so the higher share reflects a wider gap, not a shrinking baseline. The reason the baseline barely moves is that the no-platform store is operated under private forecasts and a flat tariff, which leaves it shifting heat within the day without sight of the hosting limit or the pool. Therefore, it cannot direct pre-heating to the hours where surplus would otherwise be curtailed. The value of coordination is not an artifact of the omitted flexibility. The lossy store lowers oversupply and curtailment below the lossless one while costing more. The leaked heat must be made up, and making it up absorbs surplus that the lossless store leaves to export or curtailment. Past a half-kelvin band the heat-pump thermal rating replaces the store capacity as the limiting constraint, binding in 4.5 to 7.1\% of agent-hours, which is why the result saturates. Omitting thermal mass understates self-sufficiency by about two percentage points, overstates oversupply by about two, and understates the coordination gap.
	
	\emph{Export-side tariff:} The use-of-system tariff is applied symmetrically to import and export as a stylized network charge, and the export side is an assumption, not documented practice in every zone. Setting the export-side charge to zero leaves the $\mathrm{M3}$ curtailment essentially unchanged, within 1~MWh in every deep configuration, and raises the coordination gap from \euro{}1.8k to \euro{}4.9k to \euro{}2.5k to \euro{}5.7k. A higher effective export price intensifies the competition of uncoordinated prosumers for scarce hosting capacity. The presented surplus-side results are conservative under the symmetric tariff, not an artifact of charging exports.
	
	Overall, the sensitivities show that the qualitative conclusions are robust, and that the modeled coordination value is largest where the grid is moderately constrained and where export is valuable relative to the available hosting capacity, conditions common in constrained distribution networks with active feed-in. The feed-in ratio and the hosting stringency are the two calibration inputs least pinned by public data. The base case should be read as one point within these swept ranges, not as a single deterministic figure, with the presented gaps indexed by national remuneration and grid stringency.
	
	\subsection{Uncertainty, full-year robustness, and independent numerical checks}\label{subsec_res_robust}
	
\begin{figure}[htbp]
		\centering
		\includegraphics[width=0.92\textwidth]{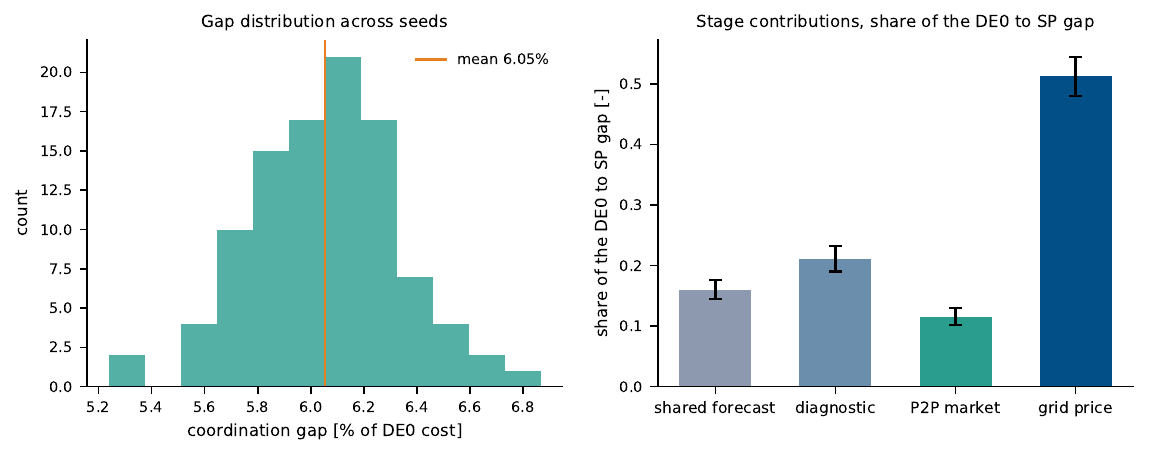}
		\caption{Robustness across 100 Monte Carlo draws for Virum, the distribution of the no-platform coordination gap on the left and the four stage contributions on the right. All four bars share one denominator, the $\mathrm{DE0}$-to-$\mathrm{SP}$ gap, and therefore sum to one. \Cref{tab_uncertainty} states the two perfect-information contributions against the $\mathrm{IND}$-to-$\mathrm{SP}$ distance instead, which is why its E2 entries are larger.}
		\label{fig_uncertainty}
	\end{figure}

	We test robustness in four complementary ways, the first and largest of which is summarized in \Cref{tab_uncertainty} and \Cref{fig_uncertainty}. First, a 100-draw experiment perturbs the weather, demand, forecast errors, day-ahead prices, and emission factors. Ambient temperature takes an additive $\mathcal N(0,1.2\,^\circ\mathrm{C})$ disturbance each hour common to all agents, solar clearness an hourly $\mathrm{Beta}(5,2)$ draw common to all agents and clipped to $[0.2,1.0]$, electrical load a multiplicative $1+\mathcal N(0,0.05)$ each hour independent per agent, the day-ahead price an additive $\mathcal N(0,0.01)$~\euro/kWh common to all agents and floored at 0.01, and the emission factor an additive $\mathcal N(0,0.01)$~kg/kWh common to all agents and floored at 0.005. The private forecast error is scaled by $1+\mathcal N(0,0.25)$ independent per agent and the shared forecast error by $1+\mathcal N(0,0.08)$ common to all agents, both clipped at 0.2 and above. Each draw uses an independent seed. The renewable portfolio is re-sized to the positive-energy target in each draw, which makes the experiment conditional on the positive-energy threshold. \Cref{fig_uncertainty} and \Cref{tab_uncertainty} show a narrow distribution of the no-platform gap and stable qualitative ordering, with the stage costs monotone from $\mathrm{DE0}$ to $\mathrm{SP}$ in every draw and tight interquartile ranges on the shared-forecast recovery, 15.1 to 17.0\% of the gap, and on the curtailment reduction at $\mathrm{M3}$, 44.0 to 47.2\%. The base run is not a typical draw. It sits at the 92nd percentile on both planner cost and shared-forecast recovery, so the 18\% quoted for the primary configuration lies above the 16.0\% mean of the distribution. A companion fixed-capacity study holds installed photovoltaics at the base value and instead varies the annual yield through a per-draw interannual weather-year multiplier (standard deviation 5\%, clipped to $\pm15$\%), letting annual generation vary through the yield. It gives a gap distribution nearly identical in mean and standard deviation to the re-sized design, with stage costs monotone and the grid price above the pool in every draw of both designs. The last column of \Cref{tab_uncertainty} compares the two designs and confirms that the coordination results are not an artifact of the re-sizing. The fixed-capacity design lets annual generation float with the weather year, and the downstream indicators nonetheless barely move. In particular, $\mathrm{M1}\!\to\!\mathrm{IND}$ remains a diagnostic residual, and the perfect-information $\mathrm{M3}$ result an upper-bound benchmark.
	
	Second, an hourly full-year-profile check tests the four-representative-week model. The latter contains 52 seven-day weeks, or 8,736 hours. It is solved as four seasonally decoupled storage problems, not as an unrestricted annual chronology. It tests the sampling of hourly profiles, not inter-seasonal storage. The full-year run gives 49.74 k\euro{} of economic cost, 0.500 hourly self-sufficiency, 0.454 oversupply, 153 MWh of curtailment, a 3.06 k\euro{} gap, and 1450 binding hours, against 50.66, 0.499, 0.455, 165, 3.17, and 1446 on the four representative weeks, relative differences of 1.8, 0.2, 0.2, 7.1, 3.5, and 0.3\%. Solving the same 8,736-hour horizon as a single annual storage cycle instead gives 49.66 k\euro, 0.501 hourly self-sufficiency, 0.453 oversupply, and 152 MWh, and that arm is a planner run alone, which is the reason it carries no gap and no binding-hour count. The coordination gap differs by 3.5\% (0.11~k\euro), of the order of its Monte Carlo standard deviation. Thus, the representative-week and full-year no-platform costs agree closely under the day-varying rolling operation with per-week storage closure. Representative weeks also bound what the model can say about storage. Four weeks cannot carry heat demand shifted across months, which matters most for the sector-coupled configurations. Relaxing that convention isolates the difference, and two installed parameters settle in advance what it can deliver. The thermal store loses 1\% of its content every hour, retaining 18\% of it after a week and almost nothing after a month, and the battery is rated at a quarter C-rate with an energy capacity near 2.6 hours of mean district load. Neither asset can hold energy from summer to winter under any horizon convention. The annual-cycle run described above, the same 8,736-hour Virum instance solved as a single storage-neutral block, confirms this. The deficit share falls by 0.13 percentage points, net import by 1.6~MWh, annual cost by \euro{}0.08k, and curtailment by 1.1~MWh. Closing more of the deficit needs seasonal storage capacity, which the model does not size, and that capacity is what the storage-investment conclusion points toward.
	
	Third, two checks confirm that the presented values are not sensitive to the numerical settings. Varying the coupling-flow regularizer from $10^{-7}$ to $10^{-5}$ leaves the $\mathrm{M3}$ economic cost, P2P and grid-price increments, pool volume, and average prices stable in qualitative terms. The same sweep applied to the forecast-informed stages, the lower block of \Cref{tab_forecast}, leaves the $\mathrm{DE0}$ and $\mathrm{M1}$ costs stable to within 0.2\% and the shared-forecast recovery within 18.2 to 19.0\% of the gap, which confirms the small forecast-informed benefit is not an artifact of the regularizer. Reversing the order of the P2P and grid-price steps gives the two-order average stated in \Cref{subsec_res_gap}. The interaction is small relative to the total gap, and the conclusion that grid-aware pricing is the larger of the two contributions at Virum is not an artifact of one waterfall order. The reversed-order check was run on Virum alone.
	
	Fourth, three further checks probe the modeling choices behind the forecast-informed step and the no-platform baseline. The upper block of \Cref{tab_forecast} varies the assumed day-ahead forecast skill. The value of the shared-forecast step $\mathrm{DE0}\!\to\!\mathrm{M1}$ scales with the gap between the private and the shared forecast error, from near zero when the private forecast is already accurate to about a third of the coordination gap when it is poor, with the base case recovering about a sixth. Strengthening the no-platform baseline has mixed effects. A 48-hour rolling horizon holds the gap at 88\% of the base value, confirming it is not an artifact of a myopic horizon. A mean-preserving time-of-use network tariff, however, lowers the gap to 56\% of the base value, where a price-responsive baseline on its own captures a substantial part of the coordination gap. The presented flat-tariff gap is an upper bound, and the value of the platform beyond simple time-of-use pricing is correspondingly smaller.

	\begin{table}[H]
		\centering
		\caption{Robustness over 100 Monte Carlo draws for Virum under both designs, the re-sized portfolio and the fixed-capacity companion, with E1 the forecast-informed experiment measured against the $\mathrm{DE0}$-to-$\mathrm{SP}$ gap, E2 the perfect-information one measured against the $\mathrm{IND}$-to-$\mathrm{SP}$ distance, and the residual between them unattributed. A dash marks an indicator the design does not produce.}
		\label{tab_uncertainty}
		\setlength{\tabcolsep}{4pt}
		\resizebox{\ifdim\width>\linewidth\linewidth\else\width\fi}{!}{%
			\begin{tabular}{@{}lrrr@{}}
				\toprule
				Indicator (100 draws) & Re-sized mean $\pm$ s.d. & Re-sized IQR (p25, p75) & Fixed capacity mean $\pm$ s.d. \\
				\midrule
				Planner cost (k\euro) & 49.61 $\pm$ 0.76 & [49.10, 50.24] & -- \\
				Annual PV generation (MWh) & 1030 $\pm$ 2 & -- & 1034 $\pm$ 47 \\
				PED positivity ratio RES$/\mathcal{D}$ & 1.05 $\pm$ 0.00 & -- & 1.05 $\pm$ 0.05 \\
				Self-sufficiency SSR & 0.499 $\pm$ 0.004 & [0.496, 0.502] & 0.499 $\pm$ 0.009 \\
				Coordination gap (k\euro) & 3.20 $\pm$ 0.14 & [3.10, 3.28] & 3.20 $\pm$ 0.14 \\
				Coordination gap (\% of DE0) & 6.05 $\pm$ 0.28 & [5.87, 6.22] & 6.07 $\pm$ 0.28 \\
				Planner curtailment (MWh) & 157.1 $\pm$ 8.7 & [152.2, 163.4] & -- \\
				Curtailment reduction at M3 (\%) & 45.7 $\pm$ 2.3 & [44.0, 47.2] & 45.7 $\pm$ 2.8 \\
				E1, shared forecast (\% of gap) & 16.0 $\pm$ 1.5 & [15.1, 17.0] & 16.0 $\pm$ 1.5 \\
				Unattributed residual, M1$\to$IND (\% of gap) & 21.1 $\pm$ 2.1 & [19.5, 22.6] & -- \\
				E2, P2P market (\% of IND$\to$SP) & 18.5 $\pm$ 2.6 & [16.8, 20.0] & -- \\
				E2, grid price (\% of IND$\to$SP) & 81.5 $\pm$ 2.6 & [80.0, 83.2] & -- \\
				Base-run percentile (cost, gap, recovery) & 92, 39, 92 & -- & -- \\
				\midrule
				Stage costs monotone, $\mathrm{DE0}$ to $\mathrm{SP}$ & 100\% of draws & -- & 100\% of draws \\
				Grid price above pool within E2 & 100\% of draws & -- & 100\% of draws \\
				\bottomrule
		\end{tabular}}
	\end{table}

	\subsection{Interpretation, settlement, and scale of the benefit}\label{subsec_res_interp}
	
	The two assumptions behind \Cref{prop_welfare} deserve a practical reading, as the communities studied hold three to six agents.  The observed integrality gap is numerically zero in every configuration, and the relaxation is exact for these instances. That is a finding about the studied instances, not a general guarantee, and a district whose battery scheduling genuinely binds the non-simultaneity constraint would retain a residue that uniform linear prices cannot implement. Price taking is the stronger assumption. A large exporter that anticipated its own effect on the congestion price could withhold export in the binding hours to lower the charge it pays, which would raise curtailment for the others and move the outcome away from $\mathrm{SP}_{\mathrm{rel}}$. The dual prices are an efficiency benchmark under that caveat and not a prediction of strategic play. Across the six configurations the largest agent holds 19 to 44\% of the binding-hour net export and the net-export Herfindahl index is 0.17 to 0.35, highest in the three-agent illustrative communities, as \Cref{tab_prices} shows. These index values lie close to the equal-share floor of $1/n$ set by the agent count, so the index reflects community size more than genuine concentration, consistent with near-symmetric binding-hour exports. That agent's congestion exposure, 8 to 16\% of its bill on a congestion payment of 1.08 to 3.09 k\euro, indicates the scale of the rent at stake.
	
	One feasible deviation tests what that exposure is worth. Capping the largest Virum exporter's net export at 90, 80 and 70\% of its base value in the hours where the limit binds, and letting the community re-optimize at the re-cleared prices, lowers its annual bill by 0.003, 0.006 and 0.010 k\euro, which is 0.02, 0.04 and 0.06\% of a 17.019 k\euro{} bill. The congestion price stays at 0.0400~\euro/kWh, community cost at 50.656 k\euro, and curtailment at 164.8 MWh at all three caps. The aggregate cap stays binding. The other agents absorb the released headroom. Across the three tested caps the congestion price that agent pays did not move. That agent forgoes \euro{}1.47k of export revenue, saves 0.82 thousand euro of congestion charge, and recovers the difference by selling into the pool. Community cost and curtailment are unchanged to three and one decimals. The tested deviations yield a negligible gain. The cap is imposed on the agent, not chosen by it, which makes that gain a lower bound on an optimized deviation, and a small lower bound places no upper bound on what an optimized deviation could extract. The finding is restricted to the three Virum deviations tested and to the congestion price observed. The remaining channel is the pool price in the hours where that agent is the marginal seller, which this check does not test. A community in which one agent's withholding could unbind the cap, or a feeder model with location-dependent limits, would give that agent price leverage. A settlement robust to market power, and strategic bidding into the pool, are left to future work.
	
	The strongest operational effect is on the surplus side. A flat tariff does not price the export-hosting externality, whereas a congestion component active in binding hours retimes local consumption, storage, and export and reduces curtailment, relative to the pool-only outcome $\mathrm{M2}$, by roughly a sixth to nearly a half across the deep configurations, and by 45\% in the primary district. This price is the shadow value of the modeled net-export cap, not a complete distribution tariff. The congestion revenue is substantial, \euro{}7k to \euro{}17k per year in the deep configurations. Across the four configurations studied in depth, annual congestion revenue was 2.9 to 3.9 times the corresponding reduction in modeled economic cost from $\mathrm{DE0}$ to $\mathrm{SP}$. Revenue recycling therefore materially affects how the gains are distributed among participants. 
	
	Without recycling, the modeled charge leaves no agent individually rational under the modeled economic-cost criterion, even though the aggregate operating cost falls. \Cref{tab_recycling} shows every prosumer's $\mathrm{M3}$ economic cost exceeding its $\mathrm{DE0}$ cost, since the unrecycled congestion payment outweighs its share of the operating saving, which places individual rationality on the settlement design. The quantity compared is an allocation of the economic objective, which carries monetized carbon and battery degradation alongside the cash terms. The table does not establish that every participant's cash bill falls. The settlement we propose is an ex ante lump sum. Each agent receives a predetermined share of the expected annual congestion revenue, \euro{}9.98k as the mean over the 100 draws. Therefore, the refund has zero derivative with respect to current behavior and the marginal congestion signal survives intact. A fixed payout preserves the marginal incentive in the modeled setting and creates a difference between realized receipts and predetermined refunds. Realized revenue in the base run is \euro{}10.50k, a surplus of \euro{}0.52k over the predetermined refunds. Funding responsibility for a shortfall in the opposite case, where realized revenue falls below the predetermined refunds, remains a settlement design decision. The reproduction package records the shortfall in each draw and assigns it to no party. Under an equal per-capita share every agent is individually rational in the base run under the modeled economic-cost criterion, though the largest exporter clears by only \euro{}0.03k and fails in 7\% of draws, by at most \euro{}182 and by \euro{}64 on average. With the predetermined PV-capacity refund, every Virum participant's modeled economic cost was at or below its $\mathrm{DE0}$ cost in every perturbation draw, since capacity share and binding-hour export share move together, which makes the rule track the agents carrying the largest payments. The modeled districts would need the proportional rule. Two ex-post rules illustrate what a refund of realized revenue would do instead. Returning the \euro{}10.5k as an equal per-capita lump sum, or in proportion to each agent's realized congestion payment, restores individual rationality in the base run, and under both the community bill after recycling is \euro{}50.66k, equal to the planner cost, since the refund returns exactly the realized revenue. Neither is implementable as a settlement, since a refund of the current period's realized payment, if anticipated, returns the very charge that creates the behavioral signal. Grid-aware pricing is an efficient allocation mechanism whose acceptability depends on a separately designed budget-balanced settlement, not a standalone tariff. A real settlement must specify how congestion revenue is recycled, how local trades are charged for network use, and whether every participant is better off, questions outside the aggregate community objective.
	
	A physical separation between coordination and storage holds across the variants. Information, trading, and grid pricing reduce avoidable surplus-side inefficiency, whereas deeper deficit-side self-sufficiency requires more storage whose marginal cost rises sharply near the feasibility boundary.
	The monetary values are gross community operating benefits that exclude platform, metering, and network-reinforcement costs and the incidence of congestion settlements. The annual gap of \euro{}1.8k to \euro{}4.9k in the deep configurations is a gross economic benchmark and a scale indicator, not a platform business case, and the forecast-informed benefit under the shared-forecast step is smaller still, \euro{}0.2k to \euro{}0.6k per year, with the full gap reached only under the perfect-information stages. Both figures are reductions in modeled economic cost, which carries monetized carbon, and not cash savings, so a break-even read from them is an economic and not a cash break-even. This forecast-informed benefit is the break-even annualized platform cost a district can justify on operation alone, namely 0.58, 0.59, 0.25, and 0.23 thousand euro per year for Virum, Limerick, V\"aster\aa s, and Graz--Berlin, under the modeled shared-forecast step. These values set a demanding test for deployment. A platform whose annualized cost exceeds 0.23 to 0.59 thousand euro per district cannot be justified on the forecast-informed saving alone in the modeled districts, and one that exceeds \euro1.8k to \euro4.9k cannot be justified even against the full perfect-information gap. Districts of this scale, three to six agents and roughly one gigawatt hour of annual electricity-equivalent demand, cannot carry a platform on operating savings by themselves. A business case has to rest elsewhere, on grid-side value such as deferred reinforcement and curtailment avoided for the network operator, on aggregation across many districts that spreads the fixed cost, or on services outside the modeled scope. We attach no figure to those channels, since the model prices none of them. The resolution of the input data also bears on these values. \Cref{tab_forecast} establishes that the shared-forecast step is sensitive to the assumed forecast skill, scaling with the gap between the private and the shared forecast error. Finer metering can change forecast performance and can reveal shorter-period mismatches and trading opportunities that hourly constructed profiles do not contain. What the forecast-skill sweep cannot settle is the direction and size of that effect, since it varies an assumed error and not the resolution of the data. Establishing either would require measured series at the finer resolution, which this paper does not have. The presented shared-forecast and pool values are therefore conditional on the assumed forecast skill and on hourly constructed profiles. The stronger contribution of the model is the physical and economic diagnosis of when annual positivity remains grid-intensive and how an hourly hosting signal can improve alignment.
	
	\section{Conclusion}\label{sec_conclusion}
	
	Positive energy districts (PED) are certified on an annual energy balance. The balance conceals the hour-by-hour, bidirectional exchange that the distribution grid experiences, in which a district imports for much of the year and exports or curtails for much of the rest. Whether coordination among prosumers can close this hourly imbalance, and how much of it only new physical capacity can close, is the question that decides what a digital platform is worth to such a district. This paper quantified the residual hourly imbalance that remains once a district is annually positive, measured the coordination gap between the same community operated by a central operator and operated as a decentralized market, and valued the information and market mechanisms that a platform can supply.
	
	The district was modeled as a multi-carrier prosumer community in which electricity and heat are coupled by heat pumps and thermal storage, with batteries, shiftable demand, and a net-export hosting limit at a shared point of common coupling. The same physical system was solved as a centralized mixed-integer quadratic program and as a price-taking peer-to-peer market equilibrium. Under convexity the equilibrium implements the regularized community optimum, and the peer-to-peer and congestion prices are the dual multipliers of the pool-clearing and hosting constraints, which makes the prices shadow values of stated physical scarcity and not assumed tariffs. Two experiments decomposed the coordination gap, one forecast-informed and one under perfect information. The model was calibrated to six European district archetypes, four studied in depth and two illustrative, each sized to the positive-energy threshold, and the results were tested through a 100-draw perturbation study, a full-year hourly check, an independent convex-solver reproduction, and sensitivities on the hosting limit, the carbon price, storage, renewable penetration, demand response, the feed-in price, building thermal mass, and the export-side tariff.
	
	Three findings are important. First, one tractable model can represent the multi-carrier district, the market among its prosumers, and the hosting constraint jointly while keeping the coordination signals interpretable. The only integrality in the community problem is the battery non-simultaneity binary, its convex relaxation carries the equilibrium, and in every configuration the mixed-integer problem, its relaxation, and the grid-aware equilibrium coincide to three decimals. Second, annual positivity leaves a large and asymmetric hourly residual, and the coordination gap on top of it is modest but real. Across the four deep archetypes hourly self-sufficiency is 48--50\% and 46--53\% of renewable output is exported or curtailed. The no-platform coordination gap is 2.8--5.9\% of the no-platform economic cost. The shared-forecast step, the one stage that plans on forecasts instead of perfect information, recovers 6--18\% of that gap. Measured against the $\mathrm{IND}$-to-$\mathrm{SP}$ distance, which is the baseline of the perfect-information experiment, the pool and grid-aware pricing divide that distance between them, with grid-aware pricing the larger of the two in four of the six configurations and the pool the larger in Västerås and Türkiye. The step from $\mathrm{M1}$ to $\mathrm{IND}$ stays outside both experiments as an unattributed diagnostic residual. Third, coordination acts almost entirely on the surplus side. Grid-aware pricing cuts curtailment by a sixth to nearly a half across the deep configurations and by 45\% in the primary district, while the deficit side barely moves and the cost of raising hourly self-sufficiency climbs convexly to a feasibility wall that the installed storage sets. In the configurations tested, closing the deficit side is an investment question and not a coordination question. Two further results sharpen the picture. Adding a within-day fabric store to the building envelope raises the coordination gap at Virum from \euro{}3.17k--4.89k, where the value of coordination is not the result of the flexibility the base model omits. The sensitivity is a Virum pre-heating test and was not run on the other configurations. 
	
	The economic reading of these results is that information, internal trading, and grid-aware pricing improve the use of assets a district already owns, at low incremental cost and with no new investment, while the deficit side is a problem in storage and other long-duration flexibility. The congestion charge that delivers the surplus-side gain raises a revenue several times the reduction in modeled economic cost it delivers. Therefore, grid-aware pricing is primarily a transfer instrument, and how its revenue is recycled decides who gains. The break-even values the paper gives are upper bounds on the annualized platform, metering, and settlement cost that operation alone can justify. At the district scale studied those bounds are small, and a platform has to earn the remainder from grid-side value or from aggregation across many districts.
	
	Several implications follow for the stakeholders that would operate such a district, under the conditions the archetypes carry, namely a hosting limit that binds for a tenth to a fifth of the year, export remunerated near half the import price, and three to six agents behind one coupling point. Community operators and distribution system operators should complement the annual positive-energy target with hourly, bidirectional indicators, since every configuration meets the annual target while leaving a residual of the size this paper measures on both sides of the balance. Districts with surplus-dominated conditions and binding export limits should exploit shared information and grid-aware operation first, evaluate storage investment next, and add generation last, since further generation mainly enlarges midday oversupply. However, note that this ordering holds only for the modeled archetypes and not as a general planning rule. For regulators, a congestion component active only in the binding hours is a targeted instrument for the export-hosting externality at the modeled interface, a single aggregate cap at the coupling point, and a predetermined refund is the settlement that keeps it acceptable to the participants.
	
	Three limitations qualify these conclusions. The results are scenario-based projections from constructed archetypes under deterministic, price-taking, single-cap conditions, and the presented values are gross operating benefits that exclude platform, metering, and network-reinforcement costs. Building thermal mass enters only as a pre-heating sensitivity, and omitting it from the base model understates both hourly self-sufficiency and the coordination gap. The shared forecast applies one error trajectory across agents, which overstates cross-agent consistency and places the shared-forecast value on the high side.
	
	Future work should validate the framework against measured pilot data, replace the aggregate hosting constraint with a network-constrained and stochastic formulation, design a settlement robust to market power and strategic bidding into the pool, embed the fabric store in the base model with a comfort cost, under-heating, and measured envelope parameters, and place the operational model inside the joint investment planning of storage, generation, and network capacity. Within its stated scope, the framework offers a reproducible way to separate the residual physical hourly imbalance from the avoidable cost of weak information, missing internal trading, and unpriced export scarcity.
	
	\section*{Acknowledgments}
	
	This work was supported by the CET Partnership, the Clean Energy Transition Partnership under the 2023 Joint Call (Ref.\ CETP-FP-2023-00114), co-funded by the European Commission (Grant No.\ 101069750) and national funding agencies, including the Innovation Fund Denmark (Case No.\ 3170-00009B), SEAI Ireland, FFG Austria (Project No.\ FO999915908), TÜBİTAK, Türkiye (Grant Nos.\ 224N158 and 224N159), the Swedish Energy Agency (Contract No.\ P2024-02859), and UEFISCDI Romania.
	
	\section*{Data and code availability}
	The data inputs, optimization model, analysis scripts, and figure-generation code used in this study are available in the \href{https://github.com/AregaGetaneh/last-20-percent}{GitHub reproduction package}. The package includes instructions for reproducing the reported tables and figures. 
	
	\section*{Declaration of generative AI usage}
	During the preparation of this work, the authors used AI for language editing, structural review, and code-package quality checks. After using this tool, the authors reviewed and edited the content as needed, verified the scientific content, and take full responsibility for the content of the publication.
	
	\appendix
	
	\section{Full formulation details and linearizations}\label{app_formulation}
	
	The absolute value in the shiftable-load bound \eqref{eq_dev} is written as the two linear inequalities $d^{\mathrm{sh}}_{at}-D^{\mathrm{sh}}_{at} \le\alpha_a(D^{\mathrm{fix}}_{at}+D^{\mathrm{sh}}_{at})$ and $D^{\mathrm{sh}}_{at}-d^{\mathrm{sh}}_{at}\le\alpha_a(D^{\mathrm{fix}}_{at}+D^{\mathrm{sh}}_{at})$. The within-day store capacity is set to $\bar q^{\mathrm{sh}}_a=W^{\mathrm{sh}}\,\overline{D^{\mathrm{sh}}_a}$ with a $W^{\mathrm{sh}}=3$~h window and $\overline{D^{\mathrm{sh}}_a}$ the mean shiftable load, and the store is reset at every day boundary by \eqref{eq_shiftclose}, which shifts demand only within the day and within a bounded duration. The tie-breaking penalties $\varepsilon^{\mathrm{s}}$ and $\varepsilon^{\mathrm{d}}$ are set to $10^{-3}$ and $10^{-4}$~\euro/kWh. The coupling-flow regularizer is $\varepsilon^{\mathrm r}=10^{-6}$~\euro/(kW$^2$h), consistent with the weighted objective function $w_t\Delta t\,\varepsilon^{\mathrm r}p^2$. It makes the regularized grid and pool flows unique and eliminates offsetting round-trip pool trades. 
	The explicit upper bounds on the grid and pool flows are set to five times the peak feasible flow and are nonbinding in every computed solution, never exceeding a fifth of the imposed limit. \Cref{alg_rolling} states the daily rolling forecast-and-realize operation used by the forecast-informed variants $\mathrm{DE0}$ and $\mathrm{M1}$, including the forecast generation, the within-week state carryover, the per-week storage closure, the committed decisions, the realization, the feasibility restoration, and the cost accounting.
	
	\begin{algorithm}[H]
		\caption{Daily rolling forecast-and-realize operation ($\mathrm{DE0}$ and $\mathrm{M1}$)}\label{alg_rolling}
		\begin{algorithmic}[1]
			\Require true series, forecast-error s.d.\ $\sigma$ ($0.25$ for $\mathrm{DE0}$, $0.08$ for $\mathrm{M1}$), window $H=24$~h
			\For{each agent $a$ and representative week $\mathcal W$}
			\State reset storage $e_a$ to the reference level (half of each capacity)
			\For{each daily window $[d,d+H)$ in $\mathcal W$}
			\State draw $f^{q}_h=\max\{1+\mathcal N(0,\sigma),\,0.2\}$ for $q\in\{\mathrm{PV},\mathrm{load},\mathrm{heat}\}$, private for $\mathrm{DE0}$ and common for $\mathrm{M1}$
			\State \textbf{plan}: solve \eqref{eq_agent} on $f^{q}$ times the true series from $e_a$, record the battery schedule
			\State \textbf{realize}: re-solve \eqref{eq_agent} on the true series with the battery schedule fixed
			\State on the last window of $\mathcal W$, pin battery and thermal storage to the reference terminal in both solves
			\State carry the realized end-of-window storage into $e_a$
			\EndFor
			\EndFor
			\State aggregate the realized per-agent flows
			\State \textbf{feasibility}: throttle net export to $\bar E_t$ by curtailing dispatched PV, pro rata to exporters
			\State \textbf{cost}: evaluate the metered objective \eqref{eq_obj} on the realized flows
		\end{algorithmic}
	\end{algorithm}
	
	\FloatBarrier
	\section{Data and calibration details}\label{app_data}
	\Cref{tab_tech} states, for every parameter family, a representative value, its status, and its source, covering the technology and flexibility parameters together with the market and emission calibration per bidding zone. \Cref{tab_agents} gives the per-agent composition of the four deep configurations. The import-price levels and shapes and the zonal emission factors are calibrated to ENTSO-E Transparency Platform data \citep{entsoe}, the volumetric use-of-system tariffs to official European electricity-price statistics \citep{eurostat_elec}, and the country-specific photovoltaic yields to PVGIS climate data \citep{pvgis}. The feed-in ratios are stylized national values inspired by the export-remuneration schemes in force in 2023--24. The assumed day-ahead photovoltaic and load forecast errors are informed by \citep{antonanzas2016review,hong2016probabilistic}, the battery degradation cost by \citep{xu2018modeling}, the thermal-storage efficiency and standing loss by \citep{sarbu2018comprehensive}, and the demand-response flexibility fraction by \citep{gils2014assessment}. The seasonal weather and archetype load parameters are set to representative European values and documented in the reproduction package. The two inputs least constrained by public data are the feed-in ratio and the net-export hosting limit, both treated as scenario assumptions and examined through sensitivity analysis. The hosting limits of 130 to 260~kW are an explicit scenario assumption, stated for interpretability as roughly a third to two thirds of a nominal 400~kVA secondary distribution transformer serving a community of the modeled scale. The limit is 13 to 22\% of peak photovoltaic power and 15 to 26\% of the unconstrained peak net export, which is why it binds in the high-generation hours. It is an export limit, and the model carries no import limit, which is why the cap constrains the surplus side alone. Peak net import is 317 to 406 kW against caps of 160 to 260 kW, exceeding the cap in every deep configuration without the model limiting it.

	\begin{table}[H]
		\centering
		\caption{Per-agent composition of the four deep configurations.}
		\label{tab_agents}
		\begin{tabularx}{\textwidth}{@{}lll r >{\raggedright\arraybackslash\hspace{0pt}}X@{}}
			\toprule
			District & Agent & Archetype & Floor [m$^2$] & Technologies (kWp, kWh) \\
			\midrule
			Virum & VI-A & res block & 3200 & PV 296, HP, batt 130, TES 220, DH \\
			Virum & VI-B & res block & 3200 & PV 296, HP, batt 110, TES 180, DH \\
			Virum & VI-C & res block & 3200 & PV 296, HP, batt 60, TES 120, DH \\
			Virum & VI-D & public & 1800 & PV 166, HP, batt 29, TES 80, DH \\
			Limerick & LK-C1 & commercial & 2500 & PV 308, HP, batt 80, TES 40 \\
			Limerick & LK-C2 & commercial & 2500 & PV 308, HP, batt 40, TES 30 \\
			Limerick & LK-R1 & res block & 3200 & PV 394, HP, batt 30, TES 60 \\
			Limerick & LK-R2 & res small & 900 & PV 111, HP, batt 10, TES 25 \\
			Limerick & LK-R3 & res small & 900 & PV 111, HP, TES 20 \\
			Limerick & LK-P & public & 1800 & PV 222, HP, batt 20, TES 30 \\
			V\"aster\aa s & VA-O1 & office & 2200 & PV 328, HP, batt 60, TES 100, DH \\
			V\"aster\aa s & VA-O2 & office & 2200 & PV 328, HP, batt 40, TES 70, DH \\
			V\"aster\aa s & VA-C & commercial & 2500 & PV 373, batt 30, DH \\
			V\"aster\aa s & VA-R & res block & 3200 & PV 477, HP, batt 25, TES 50, DH \\
			Graz--Berlin & GB-1 & res block & 3200 & PV 313, HP, batt 20, TES 200, DH \\
			Graz--Berlin & GB-2 & res block & 3200 & PV 313, HP, batt 20, TES 180, DH \\
			Graz--Berlin & GB-3 & public & 1800 & PV 176, HP, batt 15, TES 150, DH \\
			Graz--Berlin & GB-4 & commercial & 2500 & PV 244, HP, batt 20, TES 160, DH \\
			\bottomrule
		\end{tabularx}
	\end{table}

	\begin{table}[H]
		\centering
		\caption{Technology, market, and emission parameters with provenance. Zone-level rows carry seven slash-separated values in the order DK2, IE, SE3, DE, AT, TR, RO.}
		\label{tab_tech}
		\setlength{\tabcolsep}{4pt}
		\begin{tabularx}{\textwidth}{@{}>{\raggedright\arraybackslash}p{4.0cm}>{\raggedright\arraybackslash}p{4.2cm}l >{\raggedright\arraybackslash\hspace{0pt}}X@{}}
			\toprule
			Parameter & Value & Status & Source \\
			\midrule
			Day-ahead price mean $\bar\lambda^{\mathrm{imp}}$ (\euro/kWh) & 0.095/\allowbreak 0.130/\allowbreak 0.070/\allowbreak 0.120/\allowbreak 0.115/\allowbreak 0.090/\allowbreak 0.100 & Public statistical & ENTSO-E zonal day-ahead ranges, 2023--24 \\
			Day-ahead price amplitude (\euro/kWh) & 0.045/\allowbreak 0.050/\allowbreak 0.035/\allowbreak 0.060/\allowbreak 0.055/\allowbreak 0.040/\allowbreak 0.045 & Public statistical & ENTSO-E zonal day-ahead ranges, 2023--24 \\
			Feed-in ratio $\lambda^{\mathrm{exp}}/\lambda^{\mathrm{imp}}$ & 0.55/\allowbreak 0.50/\allowbreak 0.55/\allowbreak 0.50/\allowbreak 0.50/\allowbreak 0.45/\allowbreak 0.50 & Assumed & Retail feed-in against import spread, 2023--24 \\
			Use-of-system tariff $\tau^{\mathrm{use}}$ (\euro/kWh) & 0.030/\allowbreak 0.035/\allowbreak 0.028/\allowbreak 0.040/\allowbreak 0.038/\allowbreak 0.030/\allowbreak 0.032 & Public statistical & DSO volumetric network tariffs, 2023--24 \\
			Grid emission factor $\overline{\mathrm{EF}}^{\mathrm{g}}$ (kg/kWh) & 0.110/\allowbreak 0.300/\allowbreak 0.030/\allowbreak 0.350/\allowbreak 0.160/\allowbreak 0.420/\allowbreak 0.290 & Public statistical & Zonal location-based factors, 2023 \\
			Grid emission amplitude (kg/kWh) & 0.060/\allowbreak 0.080/\allowbreak 0.015/\allowbreak 0.120/\allowbreak 0.070/\allowbreak 0.090/\allowbreak 0.080 & Public statistical & Zonal location-based factors, 2023 \\
			District-heating emission factor (kg/kWh) & 0.045/\allowbreak 0.150/\allowbreak 0.030/\allowbreak 0.180/\allowbreak 0.120/\allowbreak 0.200/\allowbreak 0.160 & Public statistical & Zonal location-based factors, 2023 \\
			PV specific yield (kWh/kWp/yr) & 980/\allowbreak 950/\allowbreak 950/\allowbreak 1050/\allowbreak 1080/\allowbreak 1550/\allowbreak 1300 & Public statistical & PVGIS climate ranges, TMY \\
			\midrule
			Heat-pump COP intercept and slope & 3.2 at 7\,$^\circ$C, 0.10 per $^\circ$C & Assumed & EN~14511 datasheet range \\
			Heat-pump COP bounds & [1.8, 5.5] & Assumed & EN~14511 datasheet range \\
			Battery round-trip efficiency & 0.90 & Assumed & Li-ion datasheet typical \\
			Battery state-of-charge window & [0.15, 0.85] & Assumed & Li-ion datasheet typical \\
			Battery power-to-energy ratio & 0.25 (C/4) & Assumed & Li-ion datasheet typical \\
			Battery degradation cost & 0.020 \euro/kWh cycled & Assumed & Levelized cycle-life estimate \\
			Thermal storage efficiency and standing loss & 0.95, 0.01 per h & Assumed & Water-tank TES typical \\
			Thermal storage power-to-energy ratio & 0.50 & Assumed & Water-tank TES typical \\
			Shiftable-load share, depth $\alpha_a$, window & 0.25, 0.30, 3 h & Assumed & Demand-response literature range \\
			Forecast error s.d., private and shared & 0.25, 0.08 & Assumed & Day-ahead PV and load forecast skill \\
			District-heating price & 0.085 \euro/kWh & Assumed & Retail district-heating tariff \\
			Carbon price (base) & 100 \euro/t & Assumed & Scenario value swept in \Cref{subsec_res_sens} \\
			Hosting limit (net export) & 130--260 kW & Assumed & $\approx$1/\allowbreak 3 to 2/\allowbreak 3 of a 400 kVA substation \\
			Installed capacities and floor areas & in \Cref{tab_agents} & Calibrated & Sized to the positive-energy threshold \\
			\bottomrule
		\end{tabularx}
	\end{table}
	
	\FloatBarrier

	\FloatBarrier
	\section{Additional results}\label{app_extra}
	The complete per-configuration, per-variant key performance indicators, namely cost, emissions, self-sufficiency, hourly matching rate, oversupply, and curtailment for the five decentralized variants and the planner, are provided in the reproduction package.
	The planner objective function decomposes into energy purchase, export revenue, use-of-system tariff, battery degradation, district-heating energy, and monetized carbon. Across the configurations the purchase and export terms dominate, the degradation term is small, which confirms that storage is cycled sparingly, and the carbon term is a modest share that grows with the zone emission factor, from under 4\% of cost in low-carbon Sweden to about a fifth in Ireland and the German zone. For Virum the 46.33~k\euro{} operating cost and the 4.33~k\euro{} monetized carbon give the 50.66~k\euro{} economic cost, against a 50.86~k\euro{} raw objective, a penalty share of 0.4\%. \Cref{prop_welfare} establishes the theoretical equality $\mathrm{M3}=\mathrm{SP}_{\mathrm{rel}}$. In every configuration the mixed-integer planner $\mathrm{SP}$, its convex relaxation $\mathrm{SP}_{\mathrm{rel}}$, and the decentralized grid-aware outcome $\mathrm{M3}$ coincide to four decimals, the integrality gap $\mathrm{SP}-\mathrm{SP}_{\mathrm{rel}}$ is 0.0000 k\euro, the simultaneous battery charge and discharge energy is 0.0000 MWh, and the six planner costs are those of \Cref{tab_main}. The independent regularization check is stable over two orders of magnitude in $\varepsilon^{\mathrm r}$. Sweeping it from $10^{-7}$ to $10^{-5}$ holds the $\mathrm{M3}$ economic cost within 50.656k--50.658 k\euro, the P2P increment within 0.435k--0.443 k\euro, the grid increment within 1.386k--1.499 k\euro, the pool volume within 58.3k--60.3 MWh, the all-hour arithmetic solver-dual mean $\bar\pi_{\mathrm{all}}$ within 0.0876k--0.0879~\euro/kWh, and $\bar\tau$ within 0.0065k--0.0066~\euro/kWh, which preserves the qualitative mechanism ranking. Both orders of the P2P and grid-price mechanisms and their two-order average are given in \Cref{subsec_res_gap}, where the interaction is small relative to the total perfect-information coordination value.

	\Cref{tab_forecast} presents the forecast-skill sweep and the regularizer sensitivity of the forecast-informed stages, both described in \Cref{subsec_res_robust}.
	
	\begin{table}[H]
		\centering
		\caption{Sensitivity of the shared-forecast step $\mathrm{DE0}\!\to\!\mathrm{M1}$ for Virum to the assumed day-ahead forecast skill and to the coupling-flow regularizer, where recovery is the share of the $\mathrm{DE0}$-to-$\mathrm{SP}$ gap closed. The private error is swept with the shared error held at 0.08.}
		\label{tab_forecast}
		\setlength{\tabcolsep}{4pt}
		\resizebox{\ifdim\width>\linewidth\linewidth\else\width\fi}{!}{%
			\begin{tabular}{@{}lrrrr@{}}
				\toprule
				Swept parameter & DE0 cost (k\euro) & M1 cost (k\euro) & Shared-forecast gain (k\euro) & M1 recovery (\% of gap) \\
				\midrule
				\multicolumn{5}{@{}l}{\itshape Private forecast error s.d.} \\
				0.10 & 53.29 & 53.25 & 0.04 & 1.5 \\
				0.15 & 53.41 & 53.25 & 0.17 & 6.0 \\
				0.20 & 53.60 & 53.25 & 0.35 & 12.0 \\
				0.25 & 53.82 & 53.25 & 0.58 & 18.2 \\
				0.30 & 54.05 & 53.25 & 0.80 & 23.7 \\
				0.40 & 54.56 & 53.25 & 1.31 & 33.6 \\
				\midrule
				\multicolumn{5}{@{}l}{\itshape Coupling-flow regularizer $\varepsilon^{\mathrm r}$} \\
				$10^{-7}$ & 53.824 & 53.247 & 0.577 & 18.2 \\
				$10^{-6}$ & 53.822 & 53.247 & 0.575 & 18.2 \\
				$10^{-5}$ & 53.743 & 53.158 & 0.586 & 19.0 \\
				\bottomrule
		\end{tabular}}
	\end{table}
	
	\Cref{tab_recycling} presents the congestion-revenue settlement, described in \Cref{subsec_res_interp}.
	
	\begin{table}[H]
		\centering
		\caption{Annual economic cost by participant in Virum under $\mathrm{DE0}$ and alternative congestion-revenue refund rules. Costs include monetized carbon and battery degradation. The reported percentages give the share of draws in which each participant's cost after the refund was at or below its $\mathrm{DE0}$ cost.}
		\label{tab_recycling}
		\setlength{\tabcolsep}{4pt}
		\resizebox{\ifdim\width>\linewidth\linewidth\else\width\fi}{!}{%
			\begin{tabular}{@{}lrrrrrcrc@{}}
				\toprule
				Agent & DE0 (k\euro) & M3 (k\euro) & Ex post equal (k\euro) & Ex post prop. (k\euro) & Ex ante equal (k\euro) & IR (\% draws) & Ex ante by PV (k\euro) & IR (\% draws) \\
				\midrule
				VI-A & 13.57 & 15.42 & 12.79 & 12.74 & 12.92 & 100\% & 12.62 & 100\% \\
				VI-B & 13.76 & 15.96 & 13.33 & 13.28 & 13.46 & 100\% & 13.16 & 100\% \\
				VI-C & 14.55 & 17.02 & 14.39 & 14.27 & 14.52 & 93\% & 14.22 & 100\% \\
				VI-D & 11.94 & 12.76 & 10.14 & 10.36 & 10.27 & 100\% & 11.19 & 100\% \\
				\midrule
				Community & 53.82 & 61.15 & 50.66 & 50.66 & 51.18 & -- & 51.18 & -- \\
				\bottomrule
		\end{tabular}}
	\end{table}

	\FloatBarrier

	\setlength{\bibsep}{0pt}
		
\end{document}